\documentclass[graybox]{svmult}

\usepackage{type1cm}        
\usepackage{makeidx}         
\usepackage{graphicx}        
\usepackage{multicol}        
\usepackage[bottom]{footmisc}

\usepackage{enumitem}

\usepackage{newtxtext}       %
\usepackage{newtxmath}       

\usepackage{subfig}

\usepackage{amssymb}

\usepackage{xcolor}

\newcommand{\kinectTM}{Kinect\texttrademark}
\newcommand{\kinectTMS}{\kinectTM~}

\newcommand{\ovz}{$1$vs.$\,0$}
\newcommand{\ovo}{$1$vs.$1$}

\newcommand{\ovn}{$1$vs.$\,N$}
\newcommand{\ovnstr}{$1$vs.\,N}
\newcommand{\ovv}[1]{$1$vs.${\,#1}$}
\newcommand{\Nl}{\mathcal{N}}

\usepackage{tikz}

\usetikzlibrary{arrows}
\usepackage{pgfplots}

\makeindex             

\begin{document}

\title*{High-statistics modeling of complex pedestrian avoidance
  scenarios}
\author{Alessandro Corbetta, Lars Schilders, Federico Toschi}
\institute{Alessandro Corbetta \at Department of Applied Physics,
  Eindhoven University of Technology, The Netherlands.\\
  \email{a.corbetta@tue.nl}
\and Lars Schilders \at Department of Applied Physics, Eindhoven
University of Technology, The Netherlands.
\and Federico Toschi \at Department of Applied Physics and Department
of Mathematics and Computer Science, Eindhoven University of
Technology, The Netherlands and CNR-IAC, Rome, Italy.}
%
%
\maketitle
\abstract{
  Quantitatively modeling the trajectories and behavior of pedestrians
  walking in crowds is an outstanding fundamental challenge deeply
  connected with the physics of flowing active matter, from a
  scientific point of view, and having societal applications entailing
  individual safety and comfort, from an application perspective. \\
  In this contribution, we review a pedestrian dynamics modeling
  approach, previously proposed by the authors, aimed at reproducing
  some of the statistical features of pedestrian motion. Comparing
  with high-statistics pedestrian dynamics measurements collected in
  real-life conditions (from hundreds of thousands to millions of
  trajectories), we modeled quantitatively the statistical features of
  the undisturbed motion (i.e. in absence of interactions with other
  pedestrians) as well as the avoidance dynamics triggered by a
  pedestrian incoming in the opposite direction. This was accomplished
  through (coupled) Langevin equations with potentials including
  multiple preferred velocity states and preferred paths. In this
  chapter we review this model, discussing some of its limitations, in
  view of its extension toward a more complex case: the avoidance
  dynamics of a single pedestrian walking through a crowd that is
  moving in the opposite direction. We analyze some of the challenges
  connected to this case and present extensions to the model capable
  of reproducing some features of the motion.
  }

\section{Introduction}
\label{sec:1}
Quantitatively understanding the motion of pedestrians walking in
public shared spaces is an outstanding issue of increasing societal
urgency. The scientific challenges associated with the understanding
and modeling of human dynamics share deep connections with the physics
of active matter and with fluid
dynamics~\cite{bellomo2012modeling,moussaid2009collective,
  Moussadrspb.2009.0405,Lutz}. Growing urbanization yields higher and
higher loads of users on public infrastructures such as station hubs,
airports or museums. This translates into more complex, high-density,
crowd flow conditions, and poses increasing management challenges when
it comes to ensuring individual safety and comfort. Achieving a
quantitative comprehension and developing reliable models for the
crowd motion may help, for instance, in the design of
facilities or optimize crowd management.

Many among the proposed physical models for crowds dynamics rely on
the analogy between pedestrians and active
particles~\cite{hughes2003flow}. Either at the micro-, meso- or
macro-scopic scale~\cite{bellomo2012modeling}, pedestrians are usually
represented as self-propelling particles whose dynamics is regulated
by \textit{ad-hoc} social interaction potentials
(cf. reviews~\cite{cristiani2014multiscale,helbing2001traffic}). While
many features of crowd dynamics have been qualitatively captured by
such modeling strategies (e.g., negative correlation between crowd
density and average walking velocity, intermittent behavior at
bottlenecks, formation of lanes in presence of opposing crowd
flows~\cite{seyfried2009new,helbing1995PRE}), our quantitative
understanding remains scarce, especially in comparison with other
active matter systems~\cite{RevModPhys.85.1143}.  This likely connects
with the difficulty of acquiring high-quality data with sufficient
statistical resolution to resolve the high variability exhibited by
pedestrian behavior. Such variability includes, for instance,
different choice of paths, fluctuations in velocity, rare events, as
stopping or turning around~\cite{corbetta2016fluctuations}. Underlying
a quantitative comprehension is the capability of explaining and
modeling a given pedestrian dynamics scenario, including the
variability that is measurable across many statistically independent
realization of the same scenario.

\begin{figure}[ht]
     \centering
     \subfloat[][]{\includegraphics[width=0.25\textwidth]{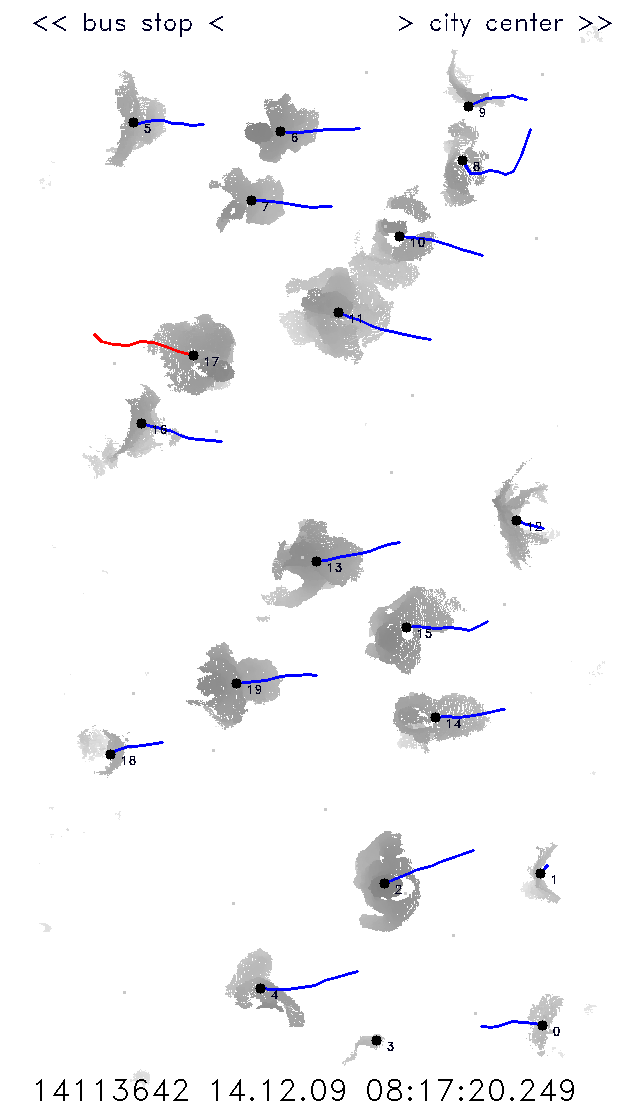}}     \subfloat[][]{\includegraphics[width=0.25\textwidth]{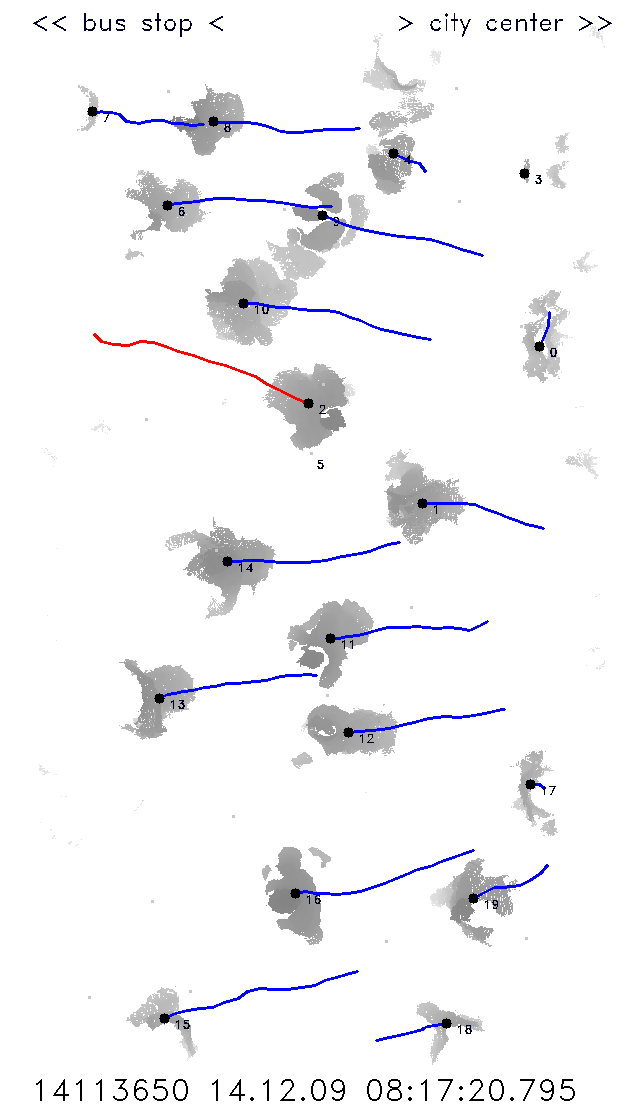}}
\subfloat[][]{\includegraphics[width=0.25\textwidth]{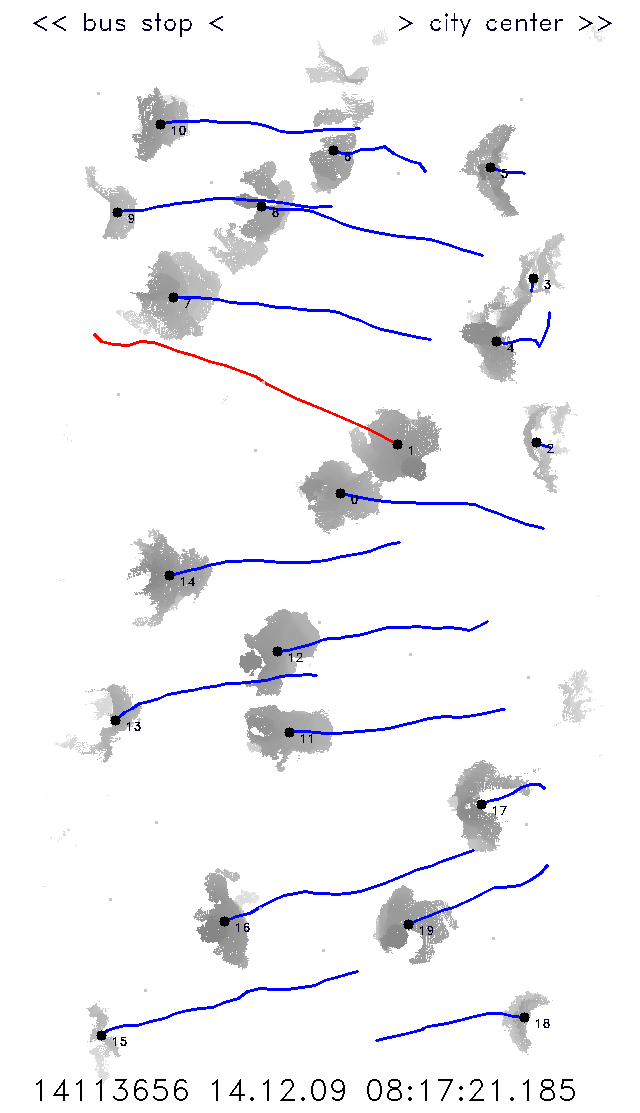}}
\subfloat[][]{\includegraphics[width=0.25\textwidth]{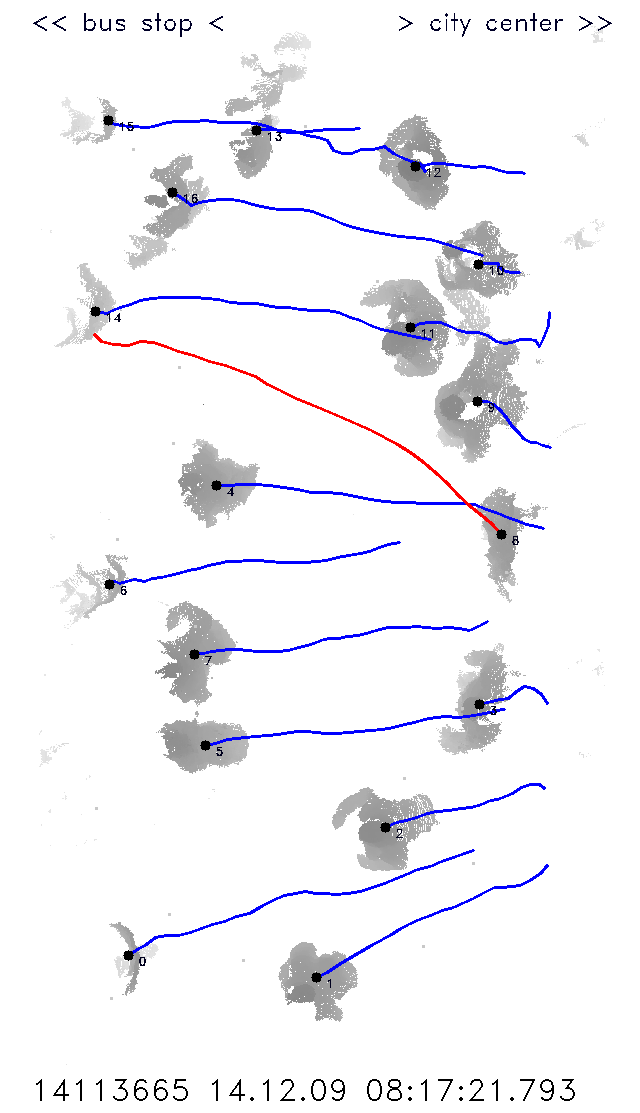}}
\caption{Avoidance scenario, \ovn, involving the target pedestrian
  walking towards the right (red trajectory) while a crowd of
  pedestrians proceeds in opposite direction (blue trajectories). We
  report four snapshots in chronological order. The target pedestrian
  escapes collisions by walking diagonally with respect to the
  longitudinal axis of the corridor (the most likely, in some sense
  ``natural'', walking direction that, in this figure, corresponds to
  the horizontal direction).  In the scenario considered, $N$ is the
  number of pedestrians walking in opposition to the target that have
  appeared in the field of view of our sensors in the time window when
  the target was present. Furthermore, we restrict to scenarios in
  which no person walking in the same direction of the target is
  present.  }
     \label{fig:snapshots_1vN}
\end{figure}

In this chapter we discuss the challenges connected to the
quantitative modeling of a relevant and ubiquitous --yet conceptually
simple-- crowd dynamics scenario which involves one pedestrian, onward
referred to as the \textit{target pedestrian}, walking in a crowd of
$N$ other pedestrians that are going in the opposite direction.  We
shall identify this scenario as \ovn, of which, in
Fig.~\ref{fig:snapshots_1vN}, we report four consecutive snapshots
taken from real-life recordings.  Our analysis employs unique
measurements collected through a months-long, 24/7, real-life
experimental campaign that targeted a section of the main walkway of
Eindhoven train station, in the Netherlands. Thanks to
state-of-the-art automated pedestrian tracking technologies, fully
developed in
house~\cite{corbetta2014TRP,corbetta2016fluctuations,%
  corbetta2016continuous,corbetta2018physics,kroneman2018accurate},
we collected millions of high-resolution pedestrian trajectories
including hundreds of occurrences of \ovn{} scenarios.

In previous works, we explored such condition in the low density
limit. In particular, we proposed quantitative models for the case of
a pedestrian walking undisturbed
(\ovz)~\cite{corbetta2016fluctuations} and for the case of a
pedestrian avoiding a single individual coming in the opposite
direction (\ovo)~\cite{corbetta2018physics}. This contribution
addresses the complexity, from the modeling and from the data analytic
points of view, arising when dealing with the \ovn{}
generalization. Our final model, assuming a superposition of pair-wise
interactions having the form proposed
in~\cite{corbetta2018physics}, involves the Newton-like dynamics
\begin{equation}
  \ddot{z}_1 = F(\dot{z}_1)+ \Nl(\,\{K(z_1,x_i), i=2,\ldots,N+1\}\,)+\sigma \dot{W},
  \label{eq:pairwise_N}
\end{equation}
where $z_1$ is a state variable including the position of the target
pedestrian, as well as their \textit{desired path} and the $\{x_i\}$'s
($i=2,\ldots,N$) are the positions of the opposing pedestrians,
$F(\dot{x_1})$ is an active term (modeling pedestrians
self-propulsion), that regulates the onward motion of the target
pedestrian, $K(x_1,x_i)$ is the pair-wise social interaction force
between the target and the $i$-th individual, $\Nl(\cdot)$ is a
(non-linear) superposition rule for the pair-wise forces. Finally, a
white Gaussian noise term $\dot{W}$, with intensity $\sigma$, provides
for stochastic fluctuations.

The present analysis shows, on the basis of high statistics
measurements, how simplifying hypotheses based on symmetry made for
the \ovz{} and \ovo{} cases
(\cite{corbetta2016fluctuations,corbetta2018physics}) do not hold in
the general \ovn{} case (as could have been expected, since the
influence of the boundary becomes relevant). Furthermore, we discuss
how the interplay of the propulsion dynamics, determined by
$F(\dot{z}_1)$, and the presence of many interaction forces,
determined by the term $\Nl(\cdot)$, may yield nonphysical effects. We
present therefore some modifications to Eq.~\eqref{eq:pairwise_N} that
enable to recover features of the observed dynamics at the
``operational level'' (e.g. local collision avoidance movements,
cf.~\cite{hoogendoorn2002normative} for a reference). This will open
the discussion on how to perform data acquisition and how to achieve
quantitative modeling to address the dynamics at the, so called,
``tactical level'', in which broader-scale individual decisions are
taken. These include, for instance, the definition of a preferred
path, selected by each individual within the current room/building, to
reach a desired destination.

This chapter is structured as follows: in Sect.~\ref{sect:setup} we
introduce our real-life pedestrian tracking setup; in
Sect.~\ref{sect:phys-ovo} we review our previous quantitative model
for pedestrians walking in diluted conditions; in
Sect.~\ref{sect:ovn-obs} we discuss through physical observables the
more generic \ovn{} scenario, and introduce some of the complexities
connected to its analysis and modeling; in Sect.~\ref{sect:ovn-model}
we address generalizations of our previous model to such case. A final
discussion in Sect.~\ref{sect:discussion} closes the chapter.


\begin{figure}[ht]
  \centering
  \includegraphics[width=.6\textwidth]{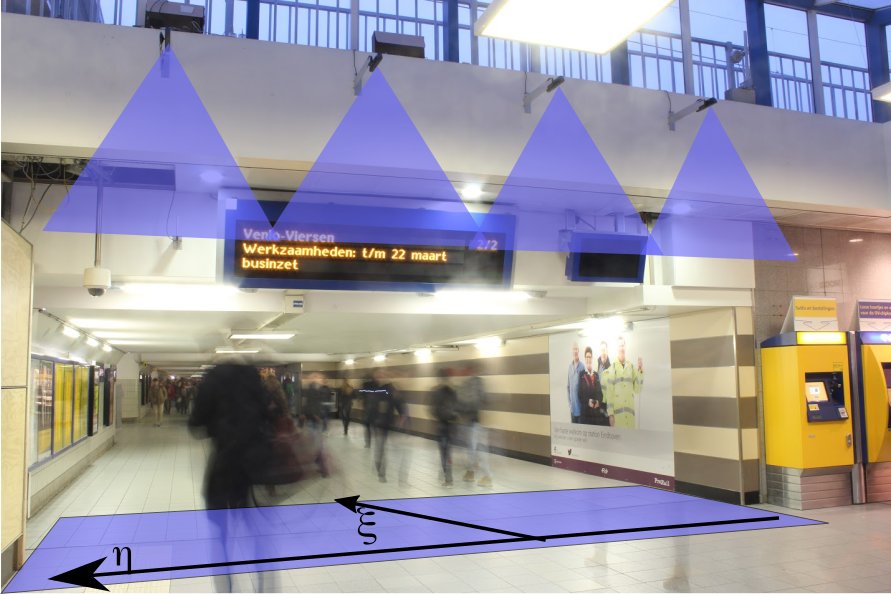}
  \caption{\label{fig:Station-measurements} Picture of the pedestrian
    tracking setup in the main walkway of Eindhoven train station
    where the data described here were collected. We overlay a sketch
    of the measurement area, reported at the floor level, and of the
    initial part of the view-cones of the four depth sensors used to
    acquire raw depth data. As the sensors had overlapping view, we
    could acquire (after appropriate data fusion) continuous depth
    maps of the full measurement area, as those in
    Fig.~\ref{fig:snapshots_1vN} (the picture is taken facing the city
    center, i.e. the observer is walking a trajectory analogous to
    that of the target pedestrian in
    Fig.~\ref{fig:snapshots_1vN}). The axes corresponding to the
    physical coordinates $(\xi,\eta)$ are also sketched (the canonical
    $(x,y)$ coordinates are here reserved for the position of a
    pedestrian in reference to their preferred path).}
\end{figure}

\section{Measurement setup and \ovnstr{} avoidance
  scenario}\label{sect:setup}
In this section, we briefly review the measurement campaign and the
technique employed to collect the data that we consider throughout
this chapter. Relevant references for the details of the campaign and
of the measurement technique are also supplied. Then we provide a
formal definition which unambiguously identifies \ovn{} scenarios.

The pedestrian dynamics data considered have been collected in the
period Oct. 2014 -- Mar.  2015 in a 24/7 pedestrian trajectory
acquisition campaign in the main walkway of Eindhoven train
station~\cite{corbetta2016continuous} (see
Fig.~\ref{fig:Station-measurements}). The measurements were collected
through a state-of-the-art pedestrian tracking system, built in-house,
and based on an array of overhead depth sensors (Microsoft
\kinectTMS~\cite{Kinect}). The sensors view-cone were in partial
overlap and allowed us to acquire data from a full transversal section
of the walkway; our observation window had a size of about $9\,m$ in
the transversal direction and of $3\,m$ in the longitudinal direction.

Depth sensors provide depth maps at a regular frame rate (in our case
$15\,$ Hz), i.e. the distance field between a point and the camera
plane. Examples of depth maps (with superimposed tracking data) are
reported in Fig.~\ref{fig:snapshots_1vN}. Notably, depth maps are
non-privacy intrusive: no features allowing individual recognition are
acquired. Nevertheless, depth maps enable accurate pedestrian
localization algorithms
(see~\cite{brscic2013person,corbetta2014TRP,seer2014kinects} for
general conceptual papers about the
technique,~\cite{corbetta2016continuous} for technical details about
this campaign, and~\cite{kroneman2018accurate} for a more recent,
highly-accurate, machine learning-based localization approach).

Our measurement location was crossed daily by several tens of
thousands people and, depending on weekday and hour, the site
underwent different crowd loads. Pedestrians could often walk
undisturbed at night hours or, more rarely, during off-peak times
(i.e. late morning and early afternoon). Else, our sensors could
measure highly variable crowding conditions ranging from
uni-directional to bi-directional flows with varying density levels.

We consider here scenarios that involve exactly one target pedestrian
walking to either of the two possible directions while other $N$
individuals are walking towards the opposite side. This means that in
accordance to our recording, the trajectory of the target pedestrian
has been perturbed exclusively by these further $N$, and no other
pedestrian walking in the direction of the target was observed
simultaneously (and thus in the
neighborhood). In~\cite{corbetta2018physics} we proposed a graph-based
approach to describe these conditions and to efficiently find them
within large databases of Lagrangian data.

\section{Physics and modeling of the diluted dynamics (\ovz{} and \ovo)}\label{sect:phys-ovo}

In this section we review the model for diluted pedestrian motion and
pairwise interactions that we proposed in our previous
papers~\cite{corbetta2016fluctuations}
and~\cite{corbetta2018physics}. We consider a crowd scenario to be
diluted whenever the target pedestrian can move freely from the
influence of other peer pedestrians (e.g. incoming, or moving close
by, i.e. \ovz{} condition) or they are just minimally affected (\ovo).

In diluted conditions, individuals crossing a corridor typically move
following (and fluctuating around) preferred paths that develop as
approximately straight trajectories. Preferred paths belong to the
tactical level of movement planning, in other words, changes in
preferred paths are connected to individual choices performed at level
overarching fine-scale navigation movements (operational level).
Without loss of generality, we consider a coordinate system such that
the state of a pedestrian can be described through three position-like
variables, $z=(x,y,y_p)$, and relative velocities (in the following
indicated, respectively, with $u$, $v$, and $\dot y_p$). In
particular, $y_p$ parameterizes the preferred path (that we assume
parallel to the $x$-axis) and $(x,y)$ identifies the instantaneous
pedestrian position.

In this reference system, as $x$ varies, individuals approach (or,
conversely, get farther apart from) their destinations. In the
transversal direction, fluctuations of amplitude $\tilde y = y-y_p$
occur around the center of the preferred path, $y_p$. In absence of
avoidance interactions with other pedestrians, we expect
$\dot y_p = 0$, at least on the tactical time-scale. Conversely, we
expect that the need of avoiding a pedestrian incoming with opposite
velocity will be reflected in a dynamics for $y_p$.

Following~\cite{corbetta2018physics}, we model the motion of a target
pedestrian in a \ovo{} condition with a Langevin dynamics as 
\begin{align}
  \frac{\mathrm{d}x}{\mathrm{d}t} &= u(t) \label{subeq3:Diff_model_int}\\
  \frac{\mathrm{d}y}{\mathrm{d}t} &= v(t) \label{subeq4:Diff_model_int}\\
  \frac{\mathrm{d}u}{\mathrm{d}t} &= - 4 \alpha_i u (u^2-u_{p,i}^2) + \sigma\dot{W}_x - e_x F_{short} \label{subeq5:Diff_model_int}\\
  \frac{\mathrm{d}v}{\mathrm{d}t} &= - 2 \lambda v - 2 \beta (y-y_{p}) + \sigma\dot{W}_y - e_y F_{short}  + F_{vision}, \label{subeq6:Diff_model_int} \\
  \frac{\mathrm{d}y_{p}}{\mathrm{d}t} &= \dot{y_{p}}(t) \label{subeq1:Diff_model_int} \\
  \frac{\mathrm{d}\dot{y}_{p}}{\mathrm{d}t} &=  F_{vision} - 2 \mu \dot{y}_{p}. \label{subeq2:Diff_model_int} 
\end{align}
In the reminder of this section we detail the expressions and the
modeling ideas underlying the preferred velocities, $u_{p,i}$, the
friction terms, $-2 \lambda v$ and $-\mu \dot y_p$, and the social
forces $F_{short}$ and $F_{vision}$. We anticipate that $F_{short}$
and $F_{vision}$ are exponentially decaying social interaction forces
depending on the distance between the target pedestrian and the other
individual. Consistently, they vanish in the case of a pedestrian
walking undisturbed thus, in such case, $\dot y_p = 0$ holds, and the
model restricts to that considered
in~\cite{corbetta2016fluctuations}. For the sake of brevity, in
Eqs.~\eqref{subeq3:Diff_model_int}-\eqref{subeq2:Diff_model_int} we
omitted the subscript ``1'' for the target pedestrian variables as in
the notation in Eq.~\eqref{eq:pairwise_N}, as in the current case
there is no ambiguity (i.e. $x$ should be written as $x_1$, and
similarly for the other variables. However, the position variables of
the second pedestrian, like $x_2$, are in fact hidden in the social
force terms).

\setlength{\tabcolsep}{12pt}
\begin{table}[th!] 
  \caption{Parameters for the model in
    Eqs.~\eqref{subeq3:Diff_model_int}-\eqref{subeq2:Diff_model_int}. 
    \label{tab:parameters}}
\centering
\begin{tabular}{l l  |  l l }
  \multicolumn{2}{c|}{\ovz{} and \ovo{}} & \multicolumn{2}{c}{\ovo{} only}                                               \\
  \hline
  \textit{Desired walking speed}         &                      & \textit{Vision f. inter. scale}     &                  \\  
  \ \ $u_{p,w}$ (walkers)                & $1.29\,$ms$^{-1}$    & \ \ $R$                             & $2.4\,$m         \\
  \textit{Desired running speed}         &                      & \textit{Contact-av f. inter. scale} &                  \\  
  \ \ $u_{p,r}$ (runners)                & $2.70\,$ms$^{-1}$    & \ \ $r$                             & $0.6\,$m         \\
  \textit{Coeff. $U(u)$, walkers}        &                      & \textit{Desired path friction}      &                  \\  
  \ \ $\alpha_w$                         & $0.037\,$m$^{-2}$s   & \ \ $\mu$                           & $1.0\,$s$^{-1}$  \\
  \textit{Coeff. $U(u)$, runners}        &                      & \textit{Vision f. intensity}        &                  \\  
  \ \ $\alpha_r$ (runners)               & $0.0015\,$m$^{-2}$s  & \ \ $A$                             & $1.5\,$ms$^{-2}$ \\
  \textit{Noise intensity}               &                      & \textit{Contact-av f. intensity}    &                  \\  
  \ \ $\sigma$                           & $0.25\,$ ms$^{-3/2}$ & \ \ $B$                             & $0.7\,$ms$^{-2}$ \\
  \textit{Transv. confinement}           &                      & \textit{Vision f. angular dep.}     &                  \\  
  \ \ $\beta$                            & $1.765\,$ m$^{-2}$s  & \ \ $\chi_{vision}$ (threshold)     & $20\,^{\circ}$   \\
  \textit{Transv. friction}              &                      & \textit{Contact-av f. angular dep.} &                  \\  
   \ \ $\lambda$                         & $0.297\,$s$^{-1}$    & \ \ $\chi_{short}$ (threshold)      & $90\,^{\circ}$   \\  
    \hline
    runner $\%$  in \ovz{}               & $4.02$\%             & runner $\%$ in  \ovo{}              & $0.2\,$\%        \\

\end{tabular}
\end{table}

\begin{figure}[ht]
  \centering
  \subfloat[][]{\includegraphics[width=0.33\textwidth]{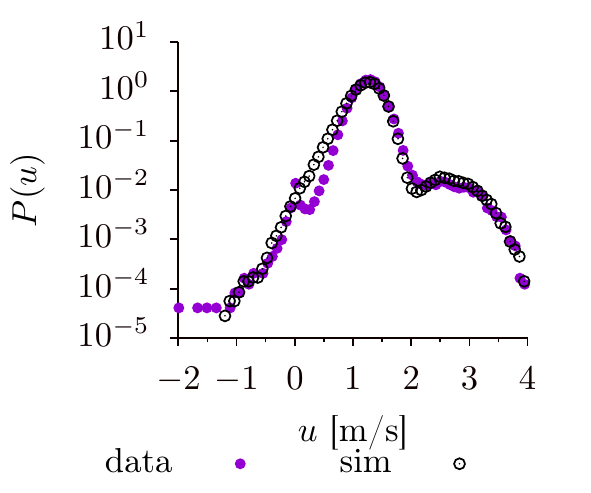}}
  \subfloat[][]{\includegraphics[width=0.33\textwidth]{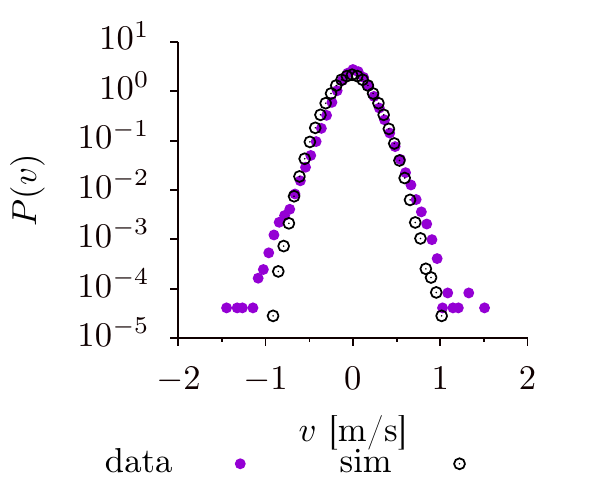}}
  \subfloat[][]{\includegraphics[width=0.33\textwidth]{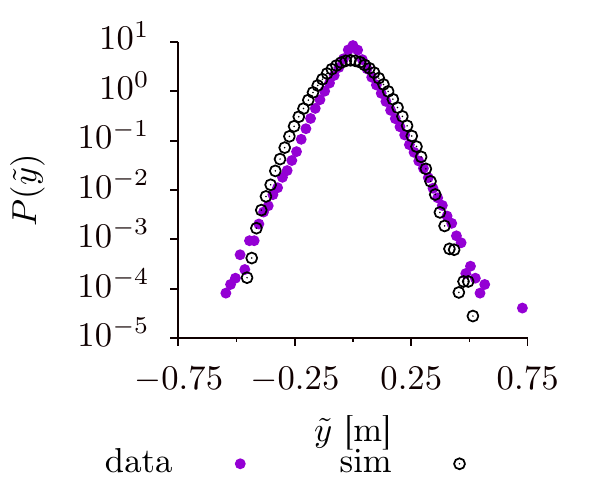}}
  \caption{\label{fig:pdf-single}Probability distribution functions of
    walking velocity and positions for pedestrians walking undisturbed
    \ovz{}: comparison between measurements (purple dots) and
    simulations of
    Eq.~\eqref{subeq3:Diff_model_int}-\eqref{subeq2:Diff_model_int} in
    absence of interaction forces (circle markers, simulation
    parameters in Tab.~\ref{tab:parameters}). The panels contain
    respectively (a) longitudinal velocities ($u$), (b) transversal
    velocities ($v$), (c) transversal positions with respect to the
    preferred path ($\tilde y = y - y_p$).  Pedestrians walk most
    frequently at around $1.29\,$m/s (cf. (a)). Besides, we observe a
    small fraction of running pedestrians, about $4\,\%$, contributing
    to the hump at above $2\,$m/s and pedestrians turning back,
    providing negative velocities contributions. (b) Transversal
    fluctuations in velocity appear to be well-approximated by a
    Gaussian distribution, while (c) transversal positions exhibit
    small deviations from a Gaussian behavior.  The model captures
    quantitatively the complete longitudinal velocity statistics
    including the running hump as well as the inversion events. The
    transversal dynamics is also well approximated as a stochastic
    damped harmonic oscillator (Eq.~\eqref{subeq4:Diff_model_int}
    and~\eqref{subeq6:Diff_model_int}).  }
\end{figure}

The second order dynamics in
Eqs.~\eqref{subeq3:Diff_model_int}-\eqref{subeq2:Diff_model_int}
includes the interplay of activity, fluctuations and
interactions. In~\cite{corbetta2016fluctuations} we showed that, in
absence of interactions, the motion of a pedestrian is characterized
by small and frequent Gaussian velocity fluctuations around a
preferred and stable velocity state, $(u,v) = (\pm u_p,0)$. Large
fluctuations can be observed as well, although rarely: for a narrow
corridor, the prominent case is the transition between the two stable
velocity states $u\rightarrow -u$, which comes with a direction
inversion. The simplest conceivable velocity potential, $U = U(u)$,
allowing for this phenomenology is a symmetric polynomial double well
with minima at $u=\pm u_p$, i.e. $U(u) \sim (u^2 - u_p^2)^2$, from
which the force term $-\partial_u U(u)\sim -u(u^2 - u_p^2)$ in
Eq.~\eqref{subeq5:Diff_model_int}. In combination with a small
Gaussian noise (term $\sigma\dot{W}_x$), this yields small-scale
Gaussian fluctuations and rare Poisson-distributed inversion events
(see also~\cite{corbetta2018path} for a path-integral based derivation
of the event statistics), both in extremely good agreement with the
data (cf.~\cite{corbetta2016fluctuations}). In
Eq.~\eqref{subeq5:Diff_model_int}, we included the subscript $i$ on
the preferred velocity and on the force intensity coefficient,
respectively $u_{p,i}$ and $\alpha_i$, to allow independent
``populations'' of pedestrians having different moving features
(e.g. walking vs. running) combined in different percentages (see
Table~\ref{tab:parameters}). In Fig.~\ref{fig:pdf-single}(a) we report
a comparison of the probability distribution function of the
longitudinal velocity $u$ for undisturbed pedestrians in case of
measurements and simulated data, which shows a remarkable agreement.

We treat the transversal dynamics as a damped stochastic harmonic
oscillator centered at $y_p$, respectively via the friction force
$-2v \lambda$, the Gaussian noise $\sigma \dot{W}_y$ and the harmonic
confinement $-2\beta (y-y_p)$. This yields Gaussian fluctuations of
$v$ and $\tilde y$, which are also in very good agreement with the
measurements, Figs.~\ref{fig:pdf-single}(b-c). For both the
longitudinal and transversal components we employ white in time
(i.e. $\delta$-correlated) and mutually uncorrelated Gaussian noise
forcing ($\dot W_x$, $\dot W_y$), with equal intensity ($\sigma$), as
validated in~\cite{corbetta2016fluctuations}. Our hypotheses on the
noise structure are guided by simplicity, yet they are somehow
arbitrary and not mandatory~\cite{Lutz}.

\begin{figure}[ht]
  \centering
\subfloat[][]{\includegraphics[width=0.45\textwidth]{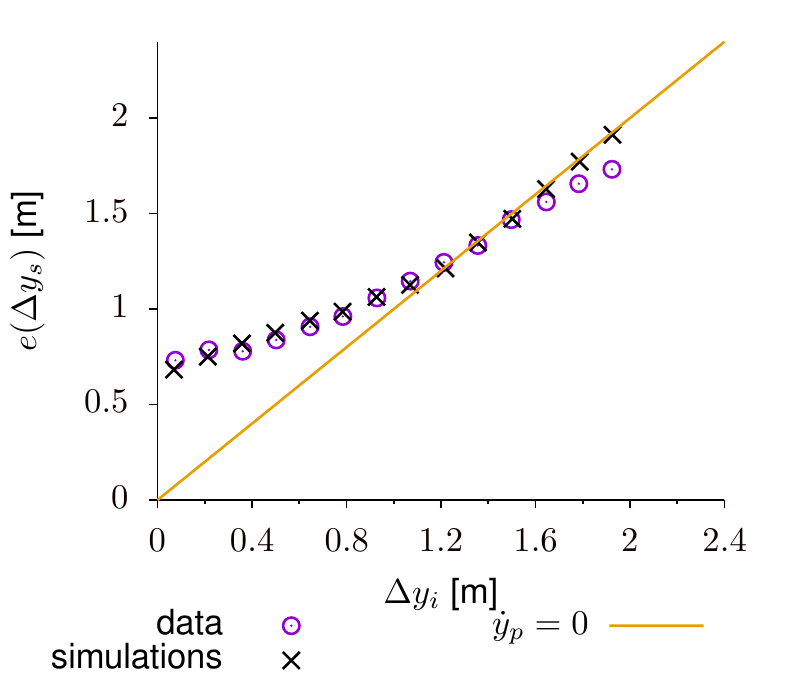}}
\subfloat[][]{\includegraphics[width=0.45\textwidth]{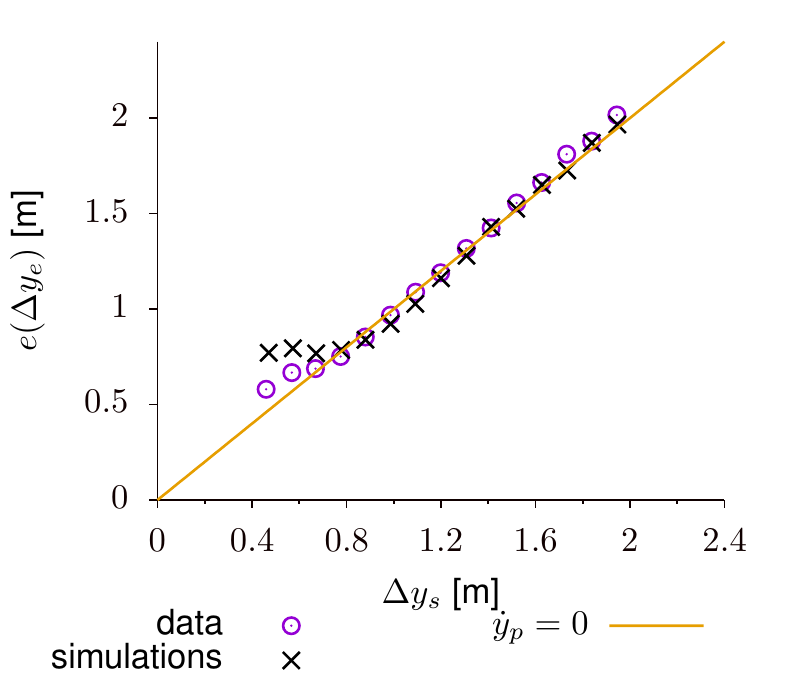}}
\caption{Average conditioned transversal distance between two
  pedestrians in a \ovo{} condition: comparison between data and
  simulations.  (a) Average, $e(|\Delta y_s|)$, of absolute lateral
  distance when at the closest point (side-by-side, $y$-axis)
  conditioned to the absolute lateral distance when at the entrance
  ($\Delta y_i$, $x$-axis).  (b) Average, $e(|\Delta y_e|)$, of
  absolute lateral distance when at the exit ($y$-axis), conditioned
  to the absolute lateral distance when side-by-side ($|\Delta y_s|$,
  x-axis). The diagonal line identifies cases in which the transversal
  distance between the pedestrians has not changed from one
  measurement point to the next, which can be interpreted as a
  preferred path that remained unchanged as pedestrian crossed the
  observation window. The model in
  Eqs.~\eqref{subeq3:Diff_model_int}-\eqref{subeq2:Diff_model_int}
  reproduce with high accuracy the avoidance
  dynamics.\label{fig:comparison-impactP}}
\end{figure}

Interactions enrich the system of social force-based coupling terms
and of a second order deterministic dynamics for $y_p$
(Eqs.~\eqref{subeq1:Diff_model_int}-\eqref{subeq2:Diff_model_int}). We
consider two conceptually different coupling forces:
\begin{itemize}
\item a long-range, vision-based, avoidance force
  \begin{equation}
    F_{vision}(x_1,y_1,x_2,y_2)= -\mathrm{sign}(e_y) A \, \mathrm{exp}(-d^2/R^2) \chi_{vision}(\tilde \theta), \label{eq:Fvisiony}
  \end{equation}
  where $e_y$ is the $y$ component of the unit vector pointing from
  $(x_1,y_1)$ to $(x_2,y_2)$ (i.e. the unit vector
  $(e_x,e_y)=(x_2-x_1,y_2-y_1)/d$, $d$ being the Euclidean distance
  between the positions of the pedestrians,
  $d = ||(x_2-x_1,y_2-y_1)||_2$), $\tilde \theta$ is the angle
  between the $x$-axis and the distance vector $(x_2-x_1,y_2-y_1)$,
  $\chi_{vision}(\tilde \theta)$ is the indicator function that is
  equal to $1$ if $|\tilde \theta| \leq 20\,^{\circ}$ and vanishing
  otherwise, $A$ and $R$ are an amplitude and a scale parameter.
\item a short-range contact-avoidance force
  \begin{equation}
    F_{short}(x_1,y_1,x_2,y_2) = B \, \mathrm{exp}(-d^2/r^2)\chi_{short}(\tilde \theta), \label{eq:Fshortx}
  \end{equation}
  where $\chi_{short}(\tilde \theta)$ is an indicator function that is
  equal to $1$ if $|\tilde \theta| \leq 90\,^{\circ}$ and vanishing otherwise,
  $B$ and $r$ are an amplitude and a scale parameter.
\end{itemize}
Note that $F_{vision}$ operates on the transversal direction only and
appears both in Eq.~\eqref{subeq6:Diff_model_int} and
Eq.~\eqref{subeq2:Diff_model_int}. In other words, it influences the
dynamics of $\tilde y$ only through $\dot{y}_p$. In fact, combining
Eqs.~\eqref{subeq6:Diff_model_int} and~\eqref{subeq2:Diff_model_int},
the evolution of $\tilde y$ satisfies
\begin{equation}
  \frac{d^2\tilde y}{dt^2} = -2\lambda\frac{d\tilde y}{dt} -2(\mu-\lambda)\dot{y}_p - 2\beta \tilde y + \sigma\dot{W}_y -e_yF_{short}.
\end{equation}
Vision and contact avoidance forces allow to reproduce the overall
avoidance dynamics. We analyze this by considering how the pedestrian
distance, projected on the $y$ direction, transversal to the motion,
changes during the avoidance maneuvers. In particular, we consider
three projected distances:
\begin{enumerate}
\item $\Delta y_i$: the absolute value of the transversal distance, as
  the pedestrians appear in our observation window;
\item $\Delta y_s$: the absolute value of the transversal distance, at
  the instant of minimum total distance between the pedestrians;
\item $\Delta y_e$: the absolute value of the transversal distance
  when the pedestrians leave our observation window.
\end{enumerate}
In Fig.~\ref{fig:comparison-impactP}, we report the conditioned
averages of these distance, comparing measurements and simulations. In
particular, Fig.~\ref{fig:comparison-impactP}(a) contains the
average transversal distance when the two pedestrians are closest
(i.e. side-by-side, $e(\Delta y_s)$), conditioned to their entrance
distance ($\Delta y_i$). We observe that for
$\Delta y_i\lessapprox 1.4\,$m avoidance maneuvers start and
pedestrians move laterally to prevent collisions. In case of
pedestrians entering facing each other ($\Delta y_i\approx 0$), on
average they establish a mutual transversal distance of about
$75\,$cm. As experience suggests, for large transversal distances no
concrete influence is measured. In
Fig.~\ref{fig:comparison-impactP}(b), we report the average
transversal distance as the two pedestrians leave the observation area
($e(\Delta y_e)$) conditioned to the transversal distance at the
moment of minimum distance ($\Delta y_s$). We observe that, on
average, the mutual distance remains unchanged. This means that the
act of avoidance impacts on the preferred path, which drifts laterally
as collision is avoided and then is not restored. We can read this as
an operational-level dynamics (avoidance gesture), that impacts on the
coarser-scale tactical-level dynamics, as the preferred path gets
changed. Remarkably, the model is capable to quantitatively recover
these features. We refer the interested reader
to~\cite{corbetta2018physics} where we additionally discuss the full
conditioned probability distributions of the transversal distances
plus other statistical observables such as pre- and post-encounter
speed and collision counts.

Note that both scenarios considered so far, \ovz{} and \ovo, feature a
translational symmetry in the transversal direction, i.e. the dynamics
is unchanged by rigid translations: $y\rightarrow y+c$,
$y_p\rightarrow y_p+c$.

\begin{figure}[ht]
  \centering    
  \subfloat[][]{\includegraphics[width=.40\textwidth,trim={2.5cm 0 2.5cm 0},clip]{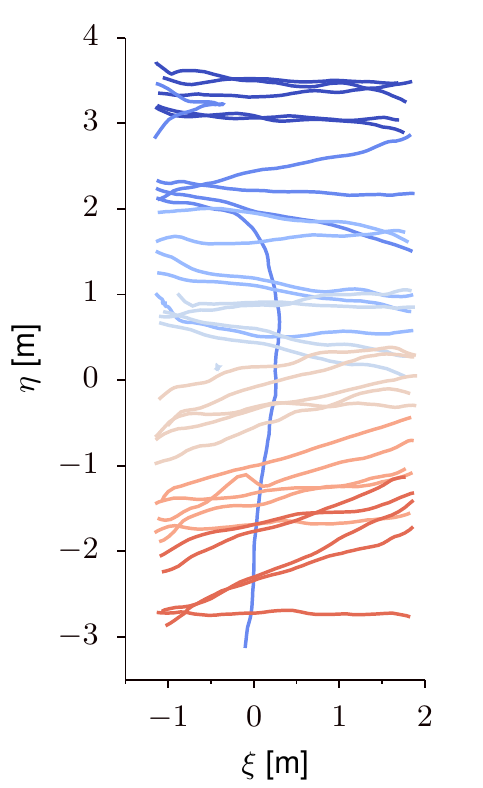}}
  \subfloat[][]{\includegraphics[width=.40\textwidth,trim={2.5cm 0 2.5cm 0},clip]{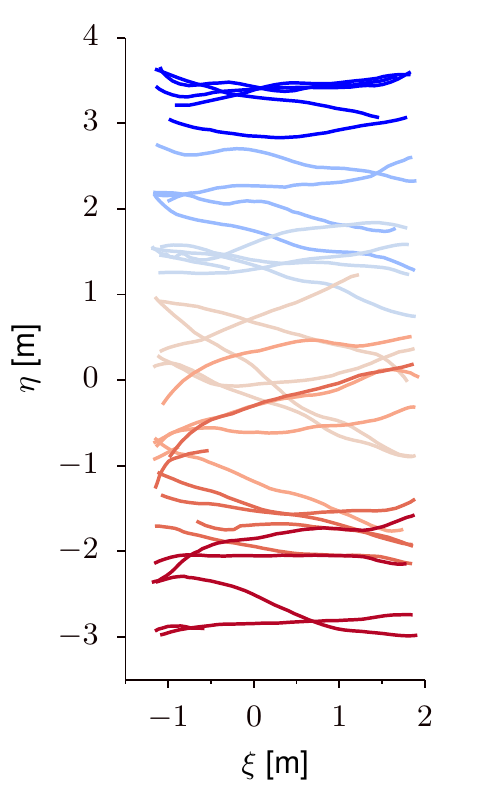}}
  \caption{\label{fig:comparison-1-vsome-traj} Random selection of
    trajectories of pedestrians walking from the bus station side of
    the train station toward the city center (from left to right in
    this reference), in case (a) of pedestrians walking alone
    (i.e. \ovz) or (b) in case \ovv{10}. In case of pedestrians
    walking alone (a), the trajectories are mostly rectilinear,
    superimposing small fluctuations to an intended path. In rare
    cases we observe large deviations, as, for instance, inversions or
    drastic trajectory changes. In case of target pedestrians facing a
    crowd (b), the trajectories exhibit ample deviations following the
    need of avoiding incoming individuals. Avoidance maneuvers
    effectively increase fluctuations, direction changes along the
    path, and dispersion in the position in which the pedestrian
    leaves the observation zone. The trajectories are here reported in
    physical coordinates $(\xi,\eta)$
    (cf. Fig.~\ref{fig:Station-measurements}).}
\end{figure}

\section{Observables of the \ovnstr{} scenario}\label{sect:ovn-obs}
As a target pedestrian walks avoiding an increasing number of other
individuals moving in the opposite direction (i.e. \ovn, $N>1$), his
or her trajectory acquires a richer and more fluctuating dynamics. In
Fig.~\ref{fig:comparison-1-vsome-traj}, we compare trajectories of
pedestrians moving towards the city center (i.e. from left to right)
in case of undisturbed pedestrians (\ovz,
Fig.~\ref{fig:comparison-1-vsome-traj}(a)) and in case \ovv{10^+}
(i.e. $N\geq 10$, Fig.~\ref{fig:comparison-1-vsome-traj}(b)). Note
that the trajectories are reported in the physical coordinate system,
$(\xi,\eta)$, where the first component is parallel to the span of the
corridor and the second component is in the transversal
direction. These coordinates must not be confused with $(x,y,y_p)$
which are instead aligned with the individual preferred paths.  In
this random sample of trajectories, it is already visible that in the
case of individual pedestrians the absence of incoming
``perturbations'' allow less pronounced fluctuations that, in most of
the cases occur around well-defined straight paths, i.e., by
definition, the preferred paths. It must be noticed that these
preferred path are not, generally, parallel to the $\xi$-axis. Rare
largely deviating trajectories also appear, in the figure it is
reported a case of trajectory inversion. Conversely, the presence of
incoming pedestrians, in addition to enhancing small-scale
fluctuations, frequently yields curved or S-like trajectories for the
target individual, as an effect of successive avoidance maneuvers.

\begin{figure}[ht]
  \centering
  \subfloat[][]{\includegraphics[width=.45\textwidth]{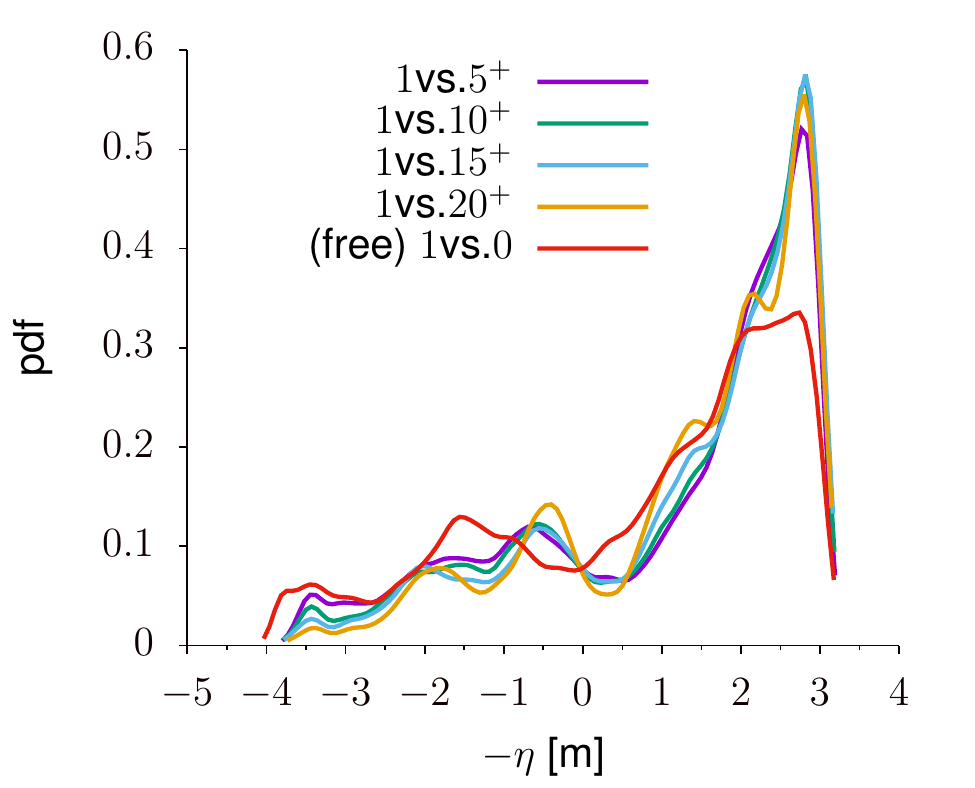}}
  \subfloat[][]{\includegraphics[width=.45\textwidth]{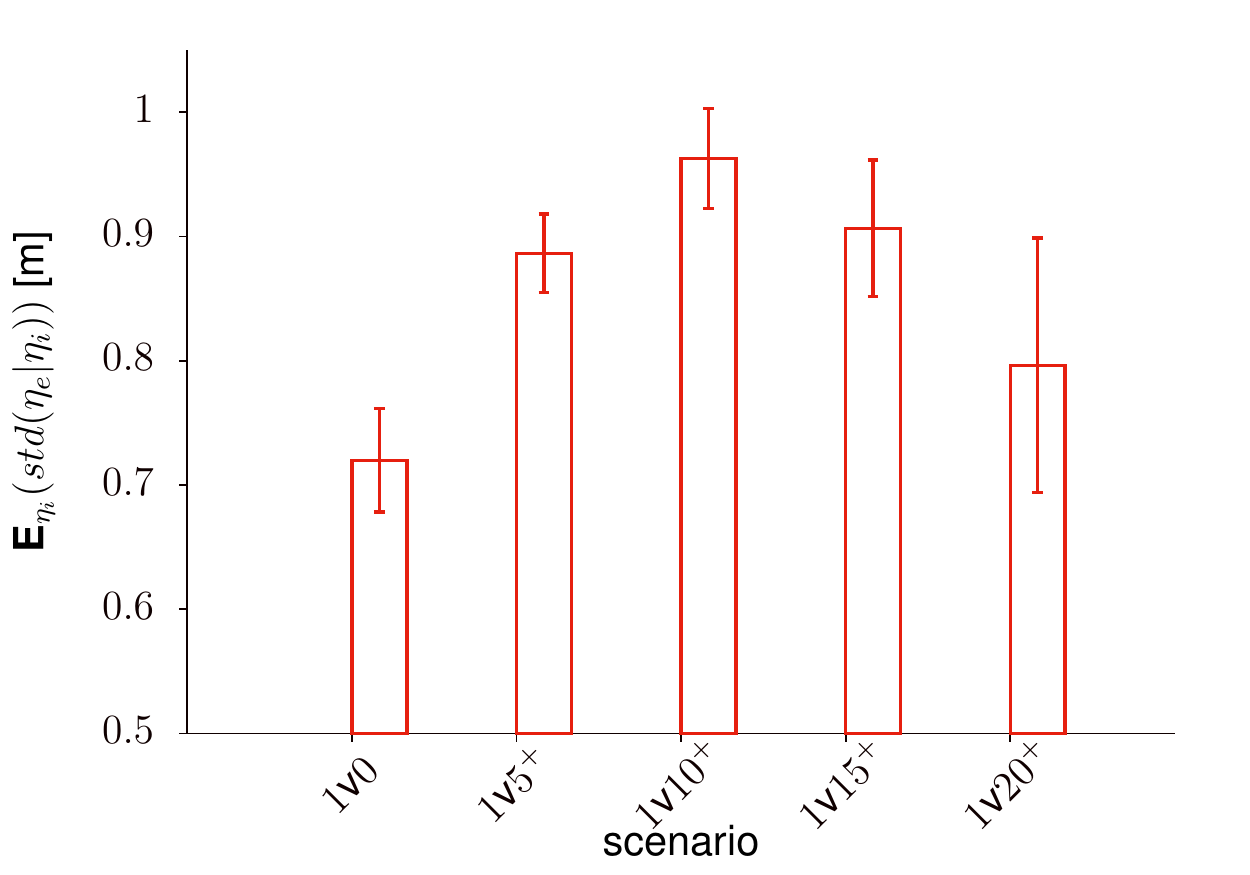}}
  \caption{(a) Probability distribution function of traversal
    positions, $\eta$, for target pedestrians in free flow (\ovz)
    vs. an increasing number of incoming pedestrians (note that the
    $\eta$ axis is flipped to $-\eta$ with respect to the reference in
    Fig.~\ref{fig:Station-measurements} such that the right side of
    the plot coincides with the right side of the corridor for an
    observer located as in Fig.~\ref{fig:Station-measurements}).
    Although the corridor is rectangular, the position distribution is
    not uniform. We believe that this possibly connects both to
    cultural biases and to the geometry upstream with respect to the
    observed areas. The entrance area is, in fact, asymmetric and
    wider on its right end. As the incoming pedestrians increase in
    number, so does the tendency to choose for the right side of the
    corridor. (b) Aggregated statistics of the outlet position
    dispersion conditioned to inlet location and incoming flow. The
    inlet distribution in (a) maps to an articulated outlet
    distribution with dependency on inlet and flow conditions. We
    report it in aggregated form by averaging the conditioned standard
    deviation of the outlet position, $std(\eta_e|\eta_i)$, over the
    inlet position $\eta_i$, (hence, no further dependency $\eta_i$
    remains. The evaluation is restricted to the bulk of the flow,
    $-1.5\,\text{m} \leq \eta_i \leq 3.1\,$m, i.e. measurements within
    half a meter from the left wall and from the right-side peak are
    neglected. Error bars report the standard error on the
    average). As the incoming crowd grows, and up to the case of $10$
    incoming pedestrians, so it grows the variance in the outlet
    position distribution. In other words, the need of avoidance
    induces larger and larger deviations from the average
    trajectory. Further increments of the number of incoming
    pedestrians yield a reduction in dispersion. This likely connects
    with the fact that the target pedestrian remains ``funneled'' by
    the incoming crowd. According to (a) this happens with highest
    probability in proximity of the right-hand side wall.
     \label{fig:stats-white-paths}
   }
\end{figure}

An incoming crowd enhances the tendency of the target pedestrian to
keep the right-hand side. In Fig.~\ref{fig:stats-white-paths}(a) we
report the probability distribution function of transversal positions
(in the corridor reference, i.e. $pdf(\eta)$). As the number of
incoming pedestrians increases, the $\eta$ distribution increasingly
peaks on the right-hand side, remaining focused in the close proximity
of the wall. Out of the bulk and close to a wall, avoidance remains
easiest and straight trajectories can be followed, see
Fig.~\ref{fig:comparison-1-vsome-traj}(b).

On the opposite, avoidance maneuvers are strongest in the bulk, and of
magnitude increasing with the number, $N$, of incoming pedestrians, at
least up to a threshold. In Fig.~\ref{fig:stats-white-paths}(b) we
report the aggregated measurement of the dispersion in the transversal
position as the target pedestrian leaves our observation window,
$\eta_e$, conditioned to their entrance position,
$\eta_i$. Specifically we report the average, computed over $\eta_i$,
of the standard deviation of $\eta_e$ conditioned to $\eta_i$, in
formulas $\textbf{E}_{\eta_i}(\,std(\eta_e|\eta_i)\,)$.  In other
words, for each entrance location (considered after a binning of the
area, $-1.5\,\text{m} \leq \eta_i \leq 3.1\,$m, into $20$ uniformly
spaced sub-regions), we consider the conditioned standard deviation of
exit location(s). As this aims at measuring the ``point-dispersion''
from each individual entrance site, we average all these
point-dispersion measurements. We notice that the average
point-dispersion increases by $20\%$ from scenario \ovz{} to \ovv{5^+}
and by a further $10\%$ when restricting to \ovv{10^+}. If we restrict
to a larger number of incoming pedestrians, the average dispersion
starts reducing. This is likely a consequence of the fact that, in
many cases, the target pedestrian remains ``funneled'' in a narrow
space left by the incoming crowd.

Considering the increment in the variability and in the fluctuations
of the trajectories for the generic \ovn{} case, for large $N$, and
the relative shortness of our observation window (about $3\,$m),
contrarily to the \ovz{} and \ovo{} cases, our data only allows us to
evaluate operational-level movements. In other words, within our
observation window we can collect statistics about the fine scale
avoidance but not on the way the preferred path gets modified on a
longer time scale. On this bases, in the next section we present a
model for the \ovn scenario.

\begin{figure}[ht]
  \centering
  \begin{tikzpicture}
    \begin{axis}
      [
      axis lines = left,
      xlabel = {$u/u_{p,\mbox{free}}\,$ [m/s]},
      ylabel = $U$,
      ymax = 7
      ]
      \addplot[
      color=blue 
      , domain=-1.5:1.5
      , samples=200
      , ultra thick
      ]{
        4 * (x^2 - 1)^2)
      };
      \addplot[
      color=red 
      , domain=-0:1.5
      , samples=200
      , ultra thick
      ]{
        16 * (x - 1)^2)
      };

      \addplot[
      color=gray 
      , domain=-0.24:1.04
      , samples=200
      , ultra thick
      , dashed
      ]{
        16 * (x - 0.4)^2)
      };
      \draw [>-<,thick] (axis cs:0.8,2)-- node[above] {$-\partial U$}
      (axis cs:1.2,2) ;
      \draw [<->,thick] (axis cs:0.8,1.1)-- node[above] {$\sigma \dot W_x$}
      (axis cs:1.2,1.1) ;

      \draw [<-,thick] (axis cs:0.8,2.7)-- node[above] {$\sim N\,F_{short}$}
      (axis cs:1.2,2.7) ;      
    \end{axis}
  \end{tikzpicture}
  \caption{\label{fig:taylor-approx} The longitudinal walking dynamics
    of a pedestrian in diluted conditions, according to
    Eqs.~\eqref{subeq3:Diff_model_int}-\eqref{subeq5:Diff_model_int},
    is defined by the interplay of a velocity gradient force,
    $-\partial U$, that brings the system toward a stable state (in
    this case, say $u=+u_p$. A sketch of the potential $U$ is reported
    in blue), a random forcing, $\sigma W_x$, that brings the system
    away from the stable state (and possibly yields transitions
    between the stable states), and the longitudinal component of the
    short-range, contact avoidance, force, $F_{short}$. As these
    forces linearly accumulate when $N$ increases, the system gets
    more and more ``unbalanced'' toward the unstable state $u=0$ or
    the negative velocities. In other words excessive forcing
    increases the hopping probability towards negative velocities and
    effectively reduces the potential barrier that separates the
    stable states. Although it is reasonable to expect that the
    probability of trajectory inversion increases in presence of a
    large incoming crowd, the phenomenon has to be probabilistically
    characterized. Here we bring this probability (unrealistically
    high for the original double-well potential, as
    in~\cite{corbetta2016fluctuations,corbetta2018physics}) to zero,
    by considering a Taylor approximation of the potential around the
    positive velocity stable state, i.e.  $U(u)\approx C (u - u_p)^p$,
    where $C$ is a positive constant. Finally, in presence of a large
    incoming crowd, the desired walking velocity (that one would keep
    in diluted flow) most likely cannot be employed due to
    ``resistance'' of the surrounding crowd. Hence, the effective
    walking velocity is reduced (cf. fundamental diagram
    in~\cite{corbetta2016continuous}). In presence of enough data the
    statistics of such reduction can be quantified. Here, focusing on
    the path fluctuations, we set the locally preferred velocity to
    the average longitudinal walking velocity of the target
    pedestrian. This translates the velocity potential towards
    velocities lower in absolute value (gray dashed line).}
\end{figure}
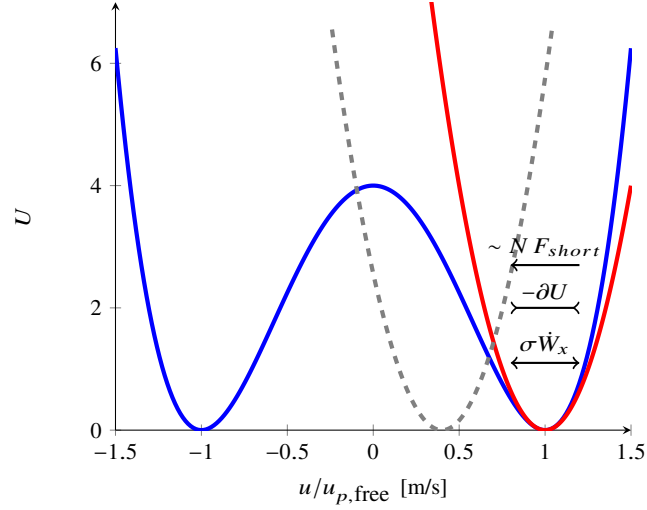

\begin{figure}[!ht]
  \centering    
  \subfloat[]{\includegraphics[width=0.25\textwidth]{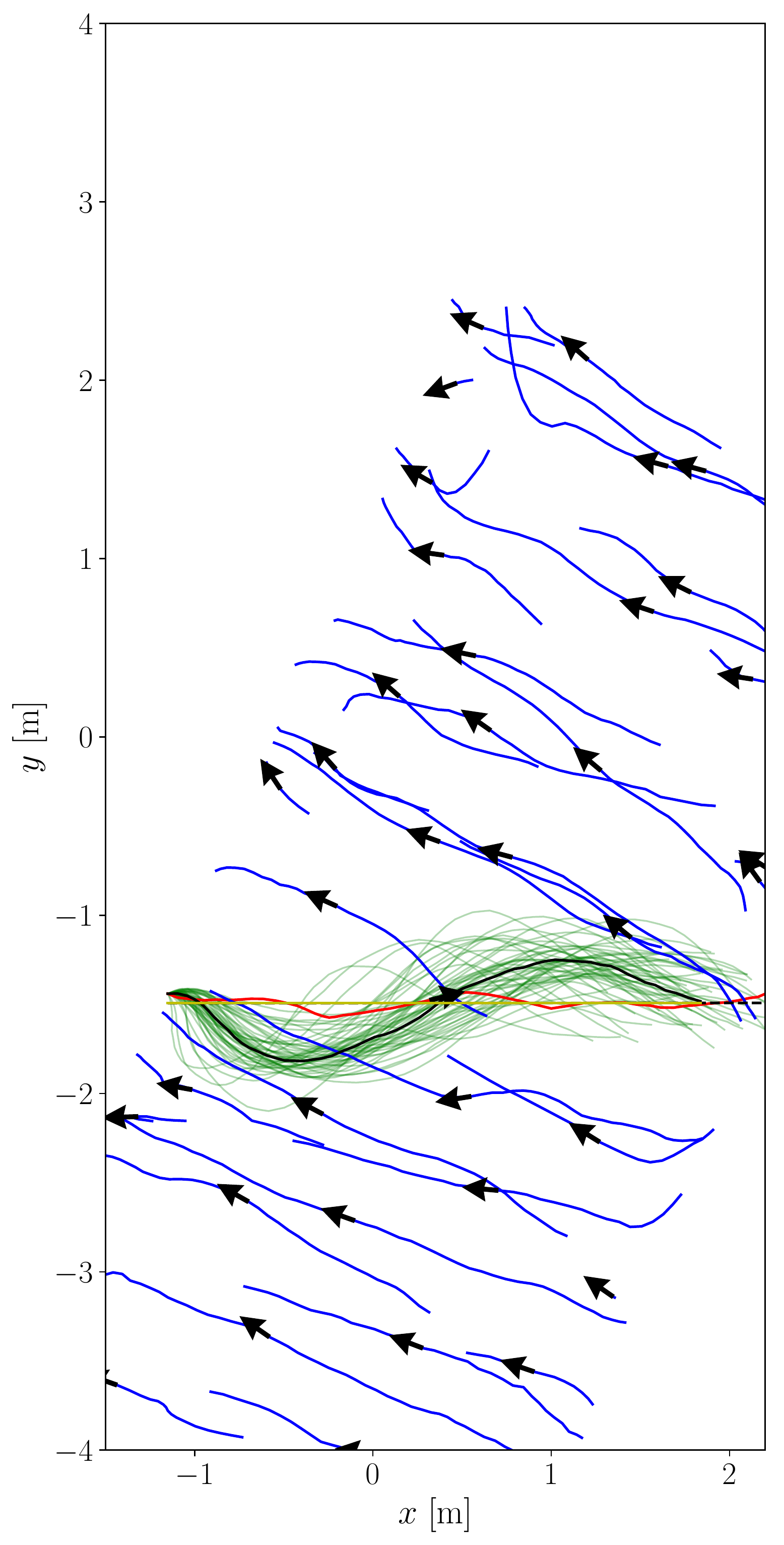}}
  \subfloat[]{\includegraphics[width=0.25\textwidth]{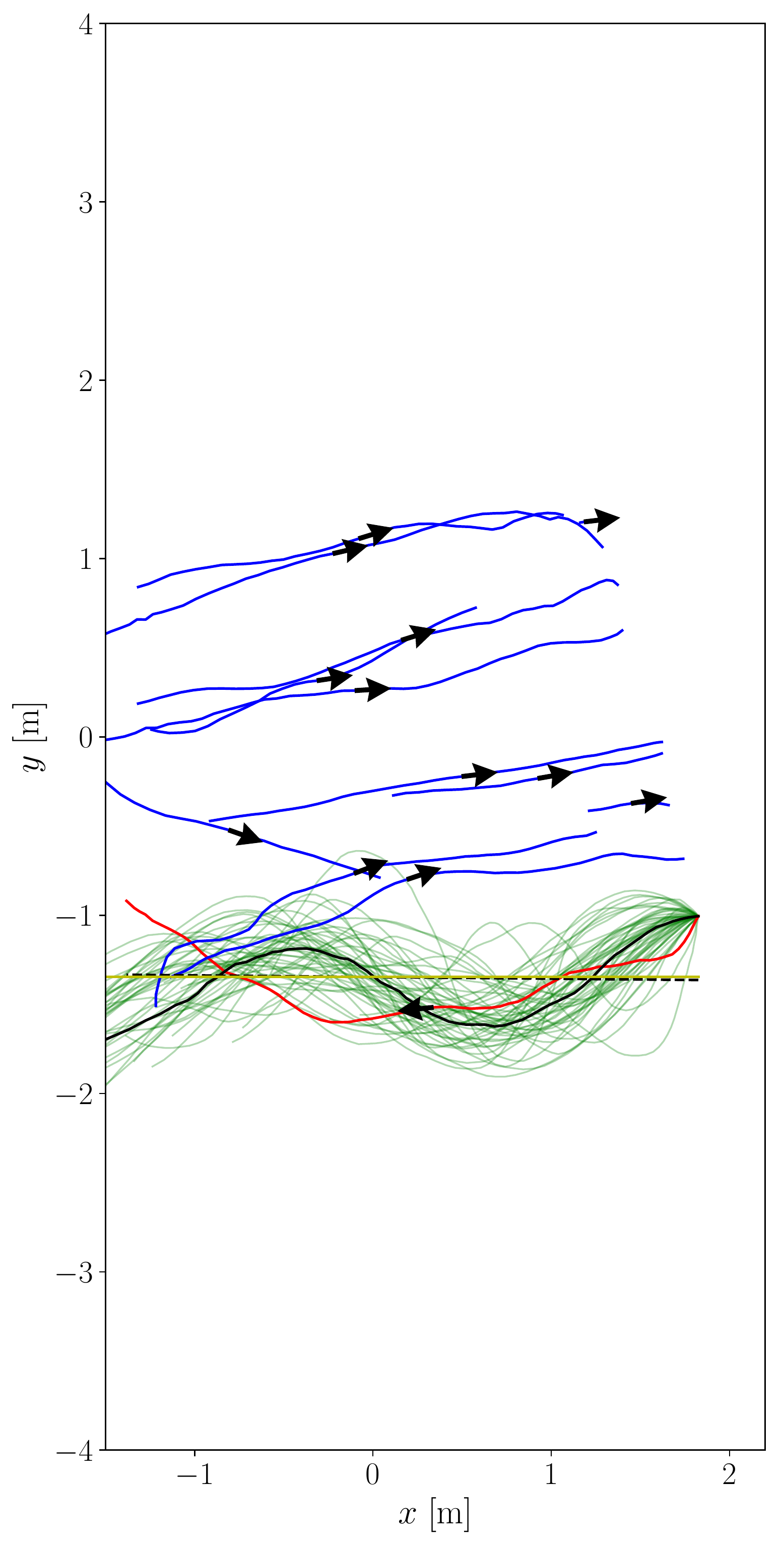}}
  \subfloat[]{\includegraphics[width=0.25\textwidth]{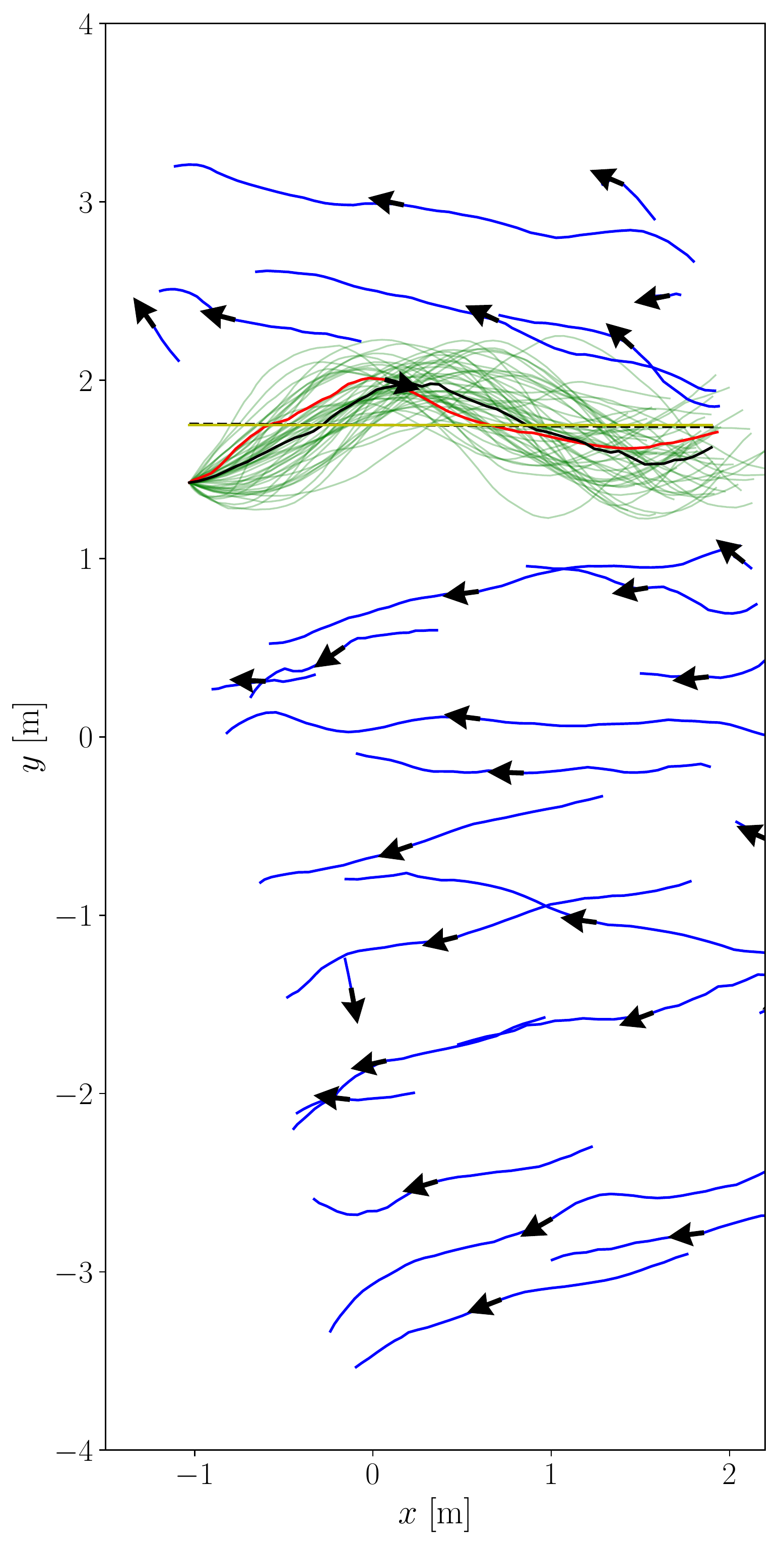}}
  \subfloat[]{\includegraphics[width=0.25\textwidth]{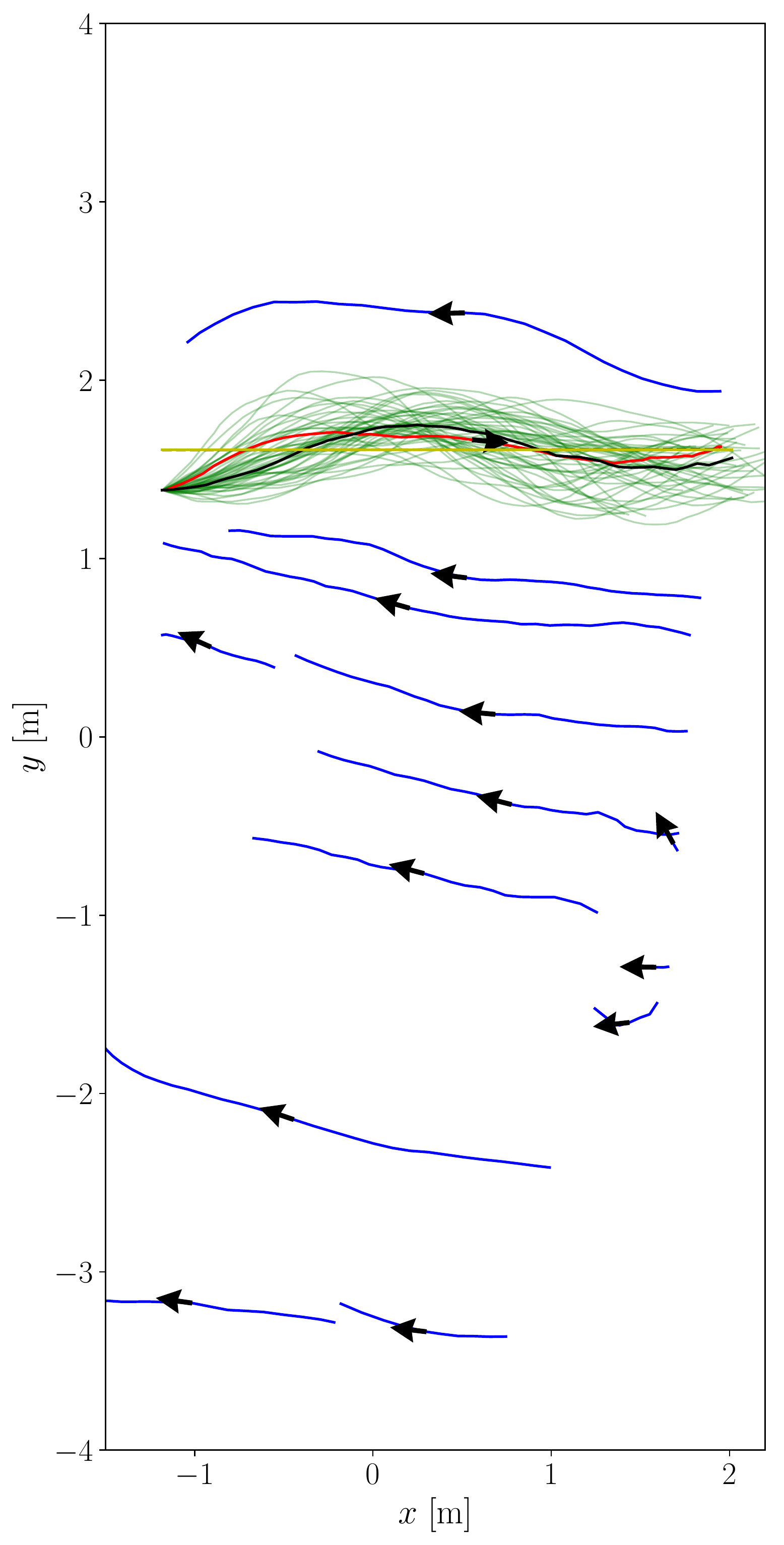}}
  \caption{Comparison of measurements and simulations performed
    via Eqs.~\eqref{eq:many-1}-\eqref{eq:many-6} and non-linear
    force superposition
    model (C4) in four \ovn{}
    scenarios. We consider the coordinates, $(x,y)$, aligned
    with the preferred path of the target pedestrian (i.e. the
    whole domain has been rotated accordingly). The measured
    trajectory of the target pedestrian is reported in red, in
    blue are the measured trajectories of the incoming crowd. We
    display in green different target pedestrian trajectory
    realizations as generated by our model, and in black 
    the classical path or the (time-)averaged simulated trajectory
    (cf. Eq.~\eqref{eq:classical-path}).          
    \label{fig:sim_taylor_aup}
  }
\end{figure}

\section{Modeling \ovnstr{} dynamics via superposition of interactions}\label{sect:ovn-model}
In this section we address the generalization of the model in
Eqs.~\eqref{subeq3:Diff_model_int}-\eqref{subeq2:Diff_model_int}
(cf. Sect.~\ref{sect:phys-ovo}) as the number of incoming pedestrians
increases.
Our underlying hypothesis is the existence of a superposition rule for
the pairwise vision-based and contact-avoidance forces in presence of
more than one opposing pedestrian. In the next equations, we indicate
these as $\Nl_{short}(\{\cdot\})$ and $\Nl_{vision}(\{\cdot\})$. To
emphasize the generality of the superposition, we set the argument of
these functions to the whole set of pairwise forces, in general
referred to as $\{f_i\}$. The linear superposition rule (or linear
superposition of effects, i.e.  $\Nl(\{f_i\}) = \sum_i f_i$),
ubiquitous in classical physics, has been widely considered in
pedestrian dynamics (e.g.,~\cite{cristiani2014BOOK,helbing1995PRE}),
but also it has been criticized
(e.g.~\cite{moussaid2011simple}). Notably, in a context of linear
superposition of forces, the total force intensity may diverge in
presence of a large crowd. Most importantly, however, it is likely
that the individual reactions are dependent on a (weighted) selection
of surrounding stimuli rather than on their blunt linear
combination~\cite{moussaid2011simple}. In formulas, we consider the
following dynamics
\begin{align}	
  & \frac{dx_1}{dt}=u_1 \label{eq:many-1}                                                                                 \\
  & \frac{du_1}{dt}=F(u_1) - \Nl_{short}(\{e_{x,i} F_{short,i}\})+\sigma_x \dot{W}_x \label{eq:many-2} \\
  & \frac{dy_1}{dt}=v_1                                                                                                   \\
  & \frac{dv_1}{dt}=-2 \lambda v_1-2 \beta(y_1-y_p)
    - \Nl_{short}(\{e_{y,i} F_{short,i}\})+ \Nl_{vision}(\{F_{vision,i}\})+\sigma_y \dot{W}_y\label{eq:pairwise_y_high}                     \\
  & \frac{dy_{p,1}}{dt}=\dot{y}_{p,1}\label{eq:pr_path_high}                                                              \\
  & \frac{d \dot{y}_{p,1}}{dt}=-2 \mu \dot{y}_{p,1} + \Nl_{vision}(\{F_{vision,i}\}),\label{eq:many-6}   
\end{align}
here the subscripts ``$1$'' and $i$ ($i=2,\ldots,N+1$) identify
explicitly the target pedestrian and the rest of the incoming
crowd. For the sake of brevity, we used the notation
$\{F_{vision,i}\}$ to indicate the set of pair-wise forces
$\{F_{vision,i}, i=2,\ldots,N+1\}$ between the target pedestrian and
the $N$ other individuals.

The highly complex dynamics, in combination with the relative
shortness of our observation window, allows us to highlight some
modeling challenges connected to finding and validating functional
forms to the terms in Eqs.~\eqref{eq:many-1}-\eqref{eq:many-6}. We
list these here and address them through additional hypotheses or
simplifications on the dynamics model.
\begin{itemize}
\item \textit{Bi-stable dynamics vs. contact avoidance forces.}
  Avoidance forces inter-play with our bi-stable velocity dynamics
  (cf. Fig.~\ref{fig:taylor-approx}). Effectively they increase the
  probability of hopping between the two stable velocity states and
  provide nonphysical trajectory inversions. Although it is reasonable
  to expect an higher trajectory inversion rates when a pedestrian
  faces a large crowd walking in opposite direction, such rate has to
  be probabilistically characterized. In modeling terms, we expect the
  height of the potential barrier $U(u_p) - U(0)$ between the stable
  velocity state, $u = \pm u_p$, and the zero walking velocity, $u=0$,
  to be altered by the incoming crowd. In absence of validation data,
  here we simplify our model by considering a second-order Taylor
  expansion of the potential $U$ around $u = +u_p$. In this way,
  $u = +u_p$ remains the only stable state of the dynamics and
  trajectory inversions are therefore impossible. While this is a
  strong simplification, it serves the present purpose of studying
  \ovn{} scenarios.
\item \textit{Preferred path.} In presence of many consecutive
  avoidance maneuvers, as in a typical \ovn{} case, the trajectory of
  the target pedestrian is continuously adjusted.  These adjustments
  likely include modifications of the preferred path. Our monitoring
  area along the longitudinal walking direction is relatively short
  (about $3\,$m).  As such, local avoidance maneuvers (operational
  level) remain mostly indistinguishable for re-adjustments of the
  preferred path (tactical level). Therefore, we opt to address path
  variations as avoidance maneuvers (i.e. operational level
  movements). As we hypothesize that tactical-level movements are
  negligible, we opt to set the preferred path to the average
  longitudinal path measured. Longer recording sites would open the
  possibility of addressing statistically the dynamics of preferred
  paths in presence of many successive interactions.
\item \textit{Preferred velocity.} The diluted motion comes with a
  measurable notion of preferred walking velocity (or velocities in
  case of multiple walking modes). In Fig.~\ref{fig:pdf-single}(a) we
  report the pdf of the longitudinal component of the velocity for
  undisturbed pedestrians, $u$, displaying the superposition of two
  dominant behaviors, pedestrians walking and running with averages
  velocity $u_{p,w}=1.29m/s$, $u_{p,r} = 2.70m/s$, respectively. In
  the generic \ovn{} case, we expect an ``adjusted'' preferred
  velocity depending on the surrounding traffic. In other words,
  although a pedestrian would keep their desired velocity constant at
  all times, the constraints given by the presence of other
  pedestrians require its temporary reduction. The velocity reduction
  is generally reported in average terms through fundamental diagrams
  (i.e. density-velocity relations~\cite{seyfried2005fundamental})
  that for our setup we quantified
  in~\cite{corbetta2016continuous}. At the microscopic level, we
  expect a number of elements influencing the adjusted preferred
  speed, e.g.: surrounding crowd density, geometry of and position in
  the domain, presence of a visible walkable free space within the
  incoming crowd, etc. These aspects are also likely statistically
  quantifiable in presence of a large enough observation window, that
  enables to disentangle tactical- and operational-level aspects of the
  dynamics. Similarly to the preferred path, here we set the preferred
  velocity to the average walking velocity of the target pedestrian.
\item \textit{Superposition rule for vision-based interactions.}
  Vision-based interactions are long-range, and relatively narrow
  angled (cf. Sect.~\ref{sect:phys-ovo}
  and~\cite{corbetta2018physics}). This makes them mostly irrelevant
  in a \ovn{} condition as in Fig.~\ref{fig:snapshots_1vN}, where
  there is limited frontal interaction as compared to interactions
  with other neighboring neighbors. As such, we opt to simplify the
  superposition rule for this forces to a linear summation, that is
  $\Nl_{vision}(\{f_i\}) = \sum_{i=2}^{N+1}f_i$.
\end{itemize}
Given these simplifications, we consider four superposition rules for
the short range contact-avoidance forces:
\begin{enumerate}[label=(C{\arabic*})]
\item $\Nl_{short}(\{f_i\}) = \sum_{i=2}^{N+1}f_i$ - this is
  a linear extension to
  Eqs.~\eqref{subeq5:Diff_model_int}-\eqref{subeq6:Diff_model_int}
  and serves as a baseline reference;
\item $\Nl_{short}(\{f_i\}) = \frac{1}{10}\sum_{i=2}^{N+1}f_i$ - this
  case is analogous (C1), but a scaling of the interaction
  by a factor $10$;
\item $\Nl_{short}(\{f_i\}) = \frac{1}{10}\sum_{i=2}^{N+1}f_i$ and
  $\alpha \rightarrow 10\alpha $ - this
  case extends (C2) by steepening the velocity potential around the
  stable velocity state by a factor $10$;
\item $\Nl_{short}(\{f_i\}) = \frac{1}{2}\max_i(f_i)$ - this is a
  non-linear superposition of forces: because of the decreasing monotonicity of the
  short-range interactions, this is equivalent to consider interactions
  exclusively with the nearest-neighbor.
\end{enumerate}
Considering the stochastic dynamics, we compare the simulations and
data as follows. We sample $200$ random occurrences in a \ovn{}
scenario from our measurement in which the target pedestrian enters in
the bulk section of the domain; each scenario is similar to what
depicted in Fig.~\ref{fig:snapshots_1vN}. We specifically consider
$N=10^+$, that according to Fig.~\ref{fig:stats-white-paths}(b), span
among the most challenging cases in terms of variability of the
paths. For each occurrence, we opt to simulate through
Eqs.~\eqref{eq:many-1}-\eqref{eq:many-6} exclusively the target
pedestrian dynamics, while we update the position of the other $N$
individuals according to the data (our simulation step, $\Delta t$, is
equal to the sampling period of the sensor,
i.e. $(15\,\mbox{Hz})^{-1}=66\,\mbox{ms}$). Employing the measured
initial position of the target pedestrian and initial velocity sampled
from the target measured walking velocities, we simulate his or her
dynamics, from the entrance in our observation window to the exit, for
$M=50$ independent realizations. We report in
Fig.~\ref{fig:sim_taylor_aup} examples of such simulations (green
lines) overlaying real measured trajectories of the target pedestrian
and of the rest of the crowd (respectively in red and blue). Employing
the simulated trajectories, we can compute an ensemble-averaged path,
$\bar z^r_1(t) = (\bar x_{1}^s(t),\bar y^s_{1}(t),\bar y^s_{p,1}(t))$
(i.e., with some abuse of terminology from the quantum path integral
language~\cite{corbetta2018path}, this would correspond to the
``classical path'') as
\begin{equation}
  \bar{z}^s_1(t) = \frac{1}{M}\sum_{k=1}^Mz^s_{1,j}(t),
  \label{eq:classical-path}
\end{equation}
where $j$ indexes the realizations, the average is performed on the
position vectors
$z_{1,j}^s(t) = (x_{1,j}^s(t),y^s_{1,j}(t),y^s_{p,1,j}(t))$ and the
superscript $s$ indicates that the quantities involved are from
simulated data.  Hence, we can compute the pdf of the instantaneous
fluctuation, $d_B(\bar z^s_1,z_1)(t)$, with respect to the classical
path
\begin{equation}
  d_B(\bar z^s_1,z_1)(t) = || (\bar x^s_1(t),\bar y^s_1(t)) - (x_1(t),y_1(t)) ||_2,
  \label{eq:dist-classic-path}
\end{equation}
where the position $z_1$ can be either from the simulated data
themselves or from the measurements. The underlying idea is to probe
how likely it is that a measured trajectory is prompted by the
model, for which, a necessary condition is a similar $d_B$ probability
distribution. Note that in the case of simulated trajectories, the
distribution of $d_B$ gives a measure for the size of the trajectory
bundle (cf. bundle of green simulated trajectories in
Fig.~\ref{fig:sim_taylor_aup}).

\begin{figure}[ht]
  \centering
  \includegraphics[width=0.99\textwidth]{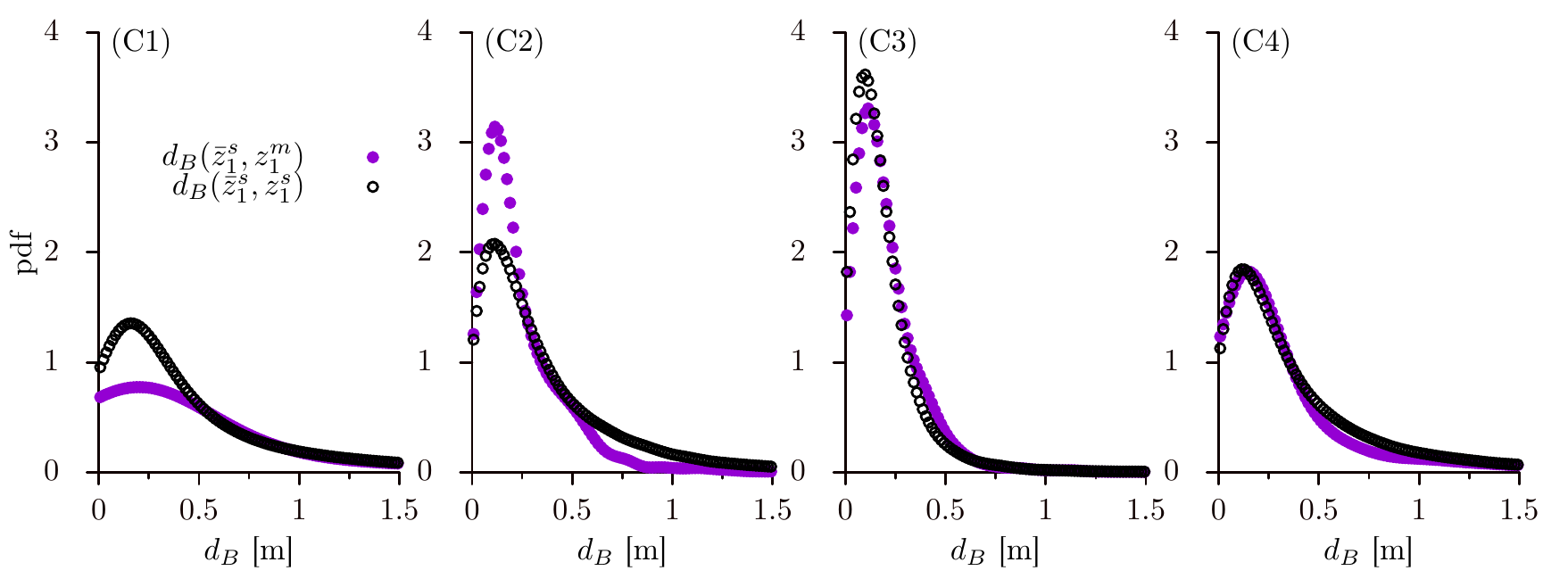}
  \caption{\label{fig:distance-comparison} Probability distribution
    functions of the distance $d_B$ between the simulated classical
    path (i.e. the average simulated trajectory,
    cf. Eq.~\eqref{eq:dist-classic-path}) and the measured or the
    simulated trajectories, respectively $d_B(\bar z^s_1,z^m_1)$ and
    $d_B(\bar z^s_1,z^s_1)$, for short-range force superposition rules
    (C1)-(C4). The non-linear force superposition, considering
    interactions with the first-neighbor only, allows highest
    similarities in the distance distributions, with minimum variation
    of parameters with respect to the \ovo{} case.}
\end{figure}

In Fig.~\ref{fig:distance-comparison}, we report the probability
distribution of the distance $d_B$ for the four force superposition
rules (C1)-(C4). We observe that a first-neighbor-only reaction (C4)
yields a distance distribution, in case of measured and simulated
trajectories, that is mutually closest while incorporating the least
parameter variations with respect to the validated \ovo{} case. In
this case, we exclusively halved the intensity of the short-range
interaction force. Such reduction might be further justified by the
fact that only the target pedestrian has been simulated, which
included no reaction of the other pedestrians that where passively
moved according to the measurements.  We stress that, possibly many
other superposition rules may exist: in case (C3) in fact, we achieved
a good agreement between the distance distribution. Nevertheless, this
involved not only a reduction of the interaction forces by a factor
$10$, which may agree with a mean-field like interaction scaling (here
$N\approx 10$ holds), but we needed to heavily steepen the velocity
potential around the stable state, with respect to the validated value
in the \ovz{} case, i.e. we increased $\alpha$ and so the likelihood
of a pedestrian to keep their desired velocity.

\section{Discussion}\label{sect:discussion}
In this chapter we addressed complex avoidance scenarios involving one
pedestrian walking in a corridor while avoiding a crowd of $N$ other
individuals walking in the opposite direction, that we conveniently
named \ovn. Our analysis has been based on real-life data collected in
an unprecedented experimental campaign held over about a one year
time-span, held in the train station of Eindhoven, The Netherlands, in
which millions of individual trajectories have been recorded with high
space- and time-resolution. We considered this scenario a first step
to tackle avoidance in non-diluted conditions; we based our analysis
and modeling on our previous works on diluted \ovz{} and \ovo{}
conditions that we briefly reviewed in the first part of the chapter.

Our contribution here is two-fold. First we evidenced, on the basis of
the experimental data, complex aspects of the dynamics arising in
comparison to a diluted flow: namely the increased randomness in the
motion, both in terms of small scale fluctuations and of avoidance
maneuvers (operational level dynamics), and the increased relevance of
geometric aspects. These elements also show how our current
trajectory database enables to explore just a small portion of the
overall \ovn{} dynamics, that we could mainly address in its
operational aspects, while we had to make assumptions on the tactical
part. 

On this basis, we considered a generalization of our previous model
for the diluted dynamics. Assuming the preferred path and speed known,
we could show that a non-linear superposition of short-ranged contact
avoidance forces, focusing on the first neighbor only, could produce a
position-wise fluctuation distribution with respect to the classical
path that was in better agreement with the measurements; i.e. with
higher chance, the trajectories measured in real-life could have been
generated by our stochastic model.  It is important to stress that
this is possibly one among many fitting forces superposition schemes.
In fact, we could produce fluctuations distributions with good
agreement between simulations and data also with a linear
superposition of forces; this however required multiple parameters
changes with respect to the validated baseline \ovz{} and \ovo{}
models.

While extending the model to the \ovn{} case we could also point out a
limitation in our \ovz{} modeling approach. We cast both types of
identified longitudinal velocity fluctuations, i.e. the frequent and
small oscillations and the rare and large path deviations (trajectory
inversions), in a unified perspective through a double well potential
in velocity. In presence of interaction forces among pedestrians,
these interplay with the gradient force due to the potential altering,
among others, the probability of inversion. Although a modification of
such probability in presence of an incoming crowd is likely, that
needs to be measured. From the modeling perspective, this modification
can be rendered in terms of a dynamic modification of the potential
barrier ($U(u_p) - U(0)$) in dependence of the surrounding crowd. This
dynamics can also be extended to other parameters of the potential,
like the preferred velocity $u_p$ that has now been inferred from the
data rather than modeled.

In general, we evidenced the increase of complexity when analyzing and
modeling dense \ovn{} avoidance scenarios vs. diluted (\ovz{} and
\ovo{}), with higher relevance of geometric aspects, mainly the
position in the domain. Moreover, in order to resolve and model
tactical level dynamics, one would require even longer measurement
campaigns, to extensively sample complex and dense pedestrian
configurations, as well as longer observation windows, to disentangle
tactical and operational level dynamics. Finally, from the modeling
perspective, we reckon that employing Langevin-like equations can get
prohibitively complex as one considers scenarios that are crowded
and/or geometrically complicated: the involved potentials, in fact,
can get excessively complex to identify and model.  On the opposite,
more trajectory-centric approaches, e.g. based on tools well
established in modern physics such as
path-integrals~\cite{corbetta2018path}, can provide more natural
modeling environments.

\begin{acknowledgement}
  We acknowledge the Brilliant Streets research program of the
  Intelligent Lighting Institute at the Eindhoven University of
  Technology, Nederlandse Spoorwegen, and the technical support of
  C. Lee, A. Muntean, T. Kanters, A. Holten, G. Oerlemans and
  M. Speldenbrink. This work is part of the JSTP research programme
  “Vision driven visitor behaviour analysis and crowd management” with
  project number 341-10-001, which is financed by the Netherlands
  Organisation for Scientific Research (NWO). A.C. acknowledges the
  support of the Talent Scheme (Veni) research programme, through
  project number 16771, which is financed by the Netherlands
  Organization for Scientific Research (NWO).
\end{acknowledgement}

\bibliographystyle{spmpsci.bst}
\bibliography{master}

\begin{thebibliography}{10}
\providecommand{\url}[1]{{#1}}
\providecommand{\urlprefix}{URL }
\expandafter\ifx\csname urlstyle\endcsname\relax
  \providecommand{\doi}[1]{DOI~\discretionary{}{}{}#1}\else
  \providecommand{\doi}{DOI~\discretionary{}{}{}\begingroup
  \urlstyle{rm}\Url}\fi

\bibitem{bellomo2012modeling}
Bellomo, N., Piccoli, B., Tosin, A.: Modeling crowd dynamics from a complex
  system viewpoint.
\newblock Mathematical Models and Methods in Applied Sciences
  \textbf{22}(supp02), 1230004 (2012)

\bibitem{brscic2013person}
Br\v{s}\v{c}i\'{c}, D., Kanda, T., Ikeda, T., Miyashita, T.: Person tracking in
  large public spaces using 3-d range sensors.
\newblock IEEE Trans. Human-Mach. Syst. \textbf{43}(6), 522--534 (2013).
\newblock \doi{10.1109/THMS.2013.2283945}

\bibitem{corbetta2014TRP}
Corbetta, A., Bruno, L., Muntean, A., Toschi, F.: High statistics measurements
  of pedestrian dynamics.
\newblock Transportation Research Procedia \textbf{2}, 96--104 (2014).
\newblock \doi{10.1016/j.trpro.2014.09.013}

\bibitem{corbetta2016fluctuations}
Corbetta, A., Lee, C., Benzi, R., Muntean, A., Toschi, F.: Fluctuations around
  mean walking behaviours in diluted pedestrian flows.
\newblock Phys. Rev. E \textbf{95}, 032316 (2017)

\bibitem{corbetta2016continuous}
Corbetta, A., Meeusen, J., Lee, C., Toschi, F.: Continuous measurements of
  real-life bidirectional pedestrian flows on a wide walkway.
\newblock In: Pedestrian and Evacuation Dynamics 2016, pp. 18--24. University
  of Science and Technology of China press (2016)

\bibitem{corbetta2018physics}
Corbetta, A., Meeusen, J.A., Lee, C.m., Benzi, R., Toschi, F.: Physics-based
  modeling and data representation of pairwise interactions among pedestrians.
\newblock Phys. Rev. E \textbf{98}(6), 062310 (2018)

\bibitem{corbetta2018path}
Corbetta, A., Toschi, F.: Path-integral representation of diluted pedestrian
  dynamics (2019)

\bibitem{cristiani2014multiscale}
Cristiani, E., Piccoli, B., Tosin, A.: Multiscale Modeling of Pedestrian
  Dynamics, vol.~12.
\newblock Springer (2014)

\bibitem{cristiani2014BOOK}
Cristiani, E., Piccoli, B., Tosin, A.: Multiscale {M}odeling of {P}edestrian
  {D}ynamics, \emph{Modeling, Simulation and Applications}, vol.~12.
\newblock Springer (2014)

\bibitem{helbing2001traffic}
Helbing, D.: Traffic and related self-driven many-particle systems.
\newblock Reviews of modern physics \textbf{73}(4), 1067 (2001)

\bibitem{helbing1995PRE}
Helbing, D., Moln\'ar, P.: Social force model for pedestrian dynamics.
\newblock Phys. Rev. E \textbf{51}(5), 4282--4286 (1995).
\newblock \doi{10.1103/PhysRevE.51.4282}

\bibitem{hoogendoorn2002normative}
Hoogendoorn, S.P., Bovy, P.H.: Normative pedestrian behaviour theory and
  modelling.
\newblock In: Transportation and Traffic Theory in the 21st Century:
  Proceedings of the 15th International Symposium on Transportation and Traffic
  Theory, Adelaide, Australia, 16-18 July 2002, pp. 219--245. Emerald Group
  Publishing Limited (2002)

\bibitem{hughes2003flow}
Hughes, R.L.: The flow of human crowds.
\newblock Annual Review of Fluid Mechanics \textbf{35}(1), 169--182 (2003)

\bibitem{kroneman2018accurate}
Kroneman, W., Corbetta, A., Toschi, F.: Accurate pedestrian localization in
  overhead depth images via height-augmented hog.
\newblock Pedestrian and Evacuation Dynamics 2018, to appear. arXiv:1805.12510
  (2018)

\bibitem{RevModPhys.85.1143}
Marchetti, M.C., Joanny, J.F., Ramaswamy, S., Liverpool, T.B., Prost, J., Rao,
  M., Simha, R.A.: Hydrodynamics of soft active matter.
\newblock Rev. Mod. Phys. \textbf{85}, 1143--1189 (2013).
\newblock \doi{10.1103/RevModPhys.85.1143}.
\newblock \urlprefix\url{https://link.aps.org/doi/10.1103/RevModPhys.85.1143}

\bibitem{Kinect}
{Microsoft Corp.}: Kinect for {X}box 360 (2012).
\newblock {R}edmond, WA, USA.

\bibitem{moussaid2009collective}
Moussa{\"\i}d, M., Garnier, S., Theraulaz, G., Helbing, D.: Collective
  information processing and pattern formation in swarms, flocks, and crowds.
\newblock Top. Cogn. Sci. \textbf{1}(3), 469--497 (2009)

\bibitem{Moussadrspb.2009.0405}
Moussa{\"\i}d, M., Helbing, D., Garnier, S., Johansson, A., Combe, M.,
  Theraulaz, G.: Experimental study of the behavioural mechanisms underlying
  self-organization in human crowds.
\newblock Proc. R. Soc. Lond., B, Biol. Sci.  (2009).
\newblock \doi{10.1098/rspb.2009.0405}

\bibitem{moussaid2011simple}
Moussa{\"\i}d, M., Helbing, D., Theraulaz, G.: How simple rules determine
  pedestrian behavior and crowd disasters.
\newblock Proceedings of the National Academy of Sciences \textbf{108}(17),
  6884--6888 (2011)

\bibitem{Lutz}
Romanczuk, P., B{\"a}r, M., Ebeling, W., Lindner, B., Schimansky-Geier, L.:
  Active {B}rownian particles.
\newblock Eur. Phys. J. Special Topics \textbf{202}(1), 1--162 (2012)

\bibitem{seer2014kinects}
Seer, S., Br{\"a}ndle, N., Ratti, C.: Kinects and human kinetics: A new
  approach for studying pedestrian behavior.
\newblock Transport. Res. C-Emer. \textbf{48}, 212--228 (2014).
\newblock \doi{10.1016/j.trc.2014.08.012}

\bibitem{seyfried2009new}
Seyfried, A., Passon, O., Steffen, B., Boltes, M., Rupprecht, T., Klingsch, W.:
  New insights into pedestrian flow through bottlenecks.
\newblock Transportation Science \textbf{43}(3), 395--406 (2009)

\bibitem{seyfried2005fundamental}
Seyfried, A., Steffen, B., Klingsch, W., Boltes, M.: The fundamental diagram of
  pedestrian movement revisited.
\newblock Journal of Statistical Mechanics: Theory and Experiment
  \textbf{2005}(10), P10002 (2005)

\end{thebibliography}

\end{document}